\documentclass[sigconf]{acmart}
\usepackage{dblfloatfix}
\AtBeginDocument{%
  }
\copyrightyear{2025}
\acmYear{2025}
\setcopyright{acmlicensed}
\acmConference[CHI '25]{CHI Conference on Human Factors in Computing Systems}{April 26-May 1, 2025}{Yokohama, Japan}
\acmBooktitle{CHI Conference on Human Factors in Computing Systems (CHI '25), April 26-May 1, 2025, Yokohama, Japan}
\acmDOI{10.1145/3706598.3713274}
\acmISBN{979-8-4007-1394-1/25/04}
\usepackage{xspace}
\usepackage{subcaption}
\usepackage{enumitem}
\usepackage{hyperref}
\usepackage{graphicx}
\newcommand{\website}[0]{\url{https://artmentor.github.io/}}
\newcommand{\dataset}[0]{\textsc{ArtMentor}\xspace}

\definecolor{art-realism}{rgb}{0.0, 0.1, 0.6}   %
\definecolor{art-deformation}{rgb}{0.58, 0.0, 0.83} %
\definecolor{art-imagination}{rgb}{1.0, 0.55, 0.0}  %
\definecolor{art-color-richness}{rgb}{0.0, 0.69, 0.0} %
\definecolor{art-color-contrast}{rgb}{0.69, 0.0, 0.0} %
\definecolor{art-line-combination}{rgb}{0.45, 0.45, 0.45} %
\definecolor{art-line-texture}{rgb}{0.58, 0.29, 0.0} %
\definecolor{art-picture-organization}{rgb}{0.0, 0.63, 0.63} %
\definecolor{art-transformation}{rgb}{0.0274, 0.4274, 0.2352} %
\definecolor{computer-green}{rgb}{0.2980, 0.3843, 0.2196}
\definecolor{human-orange}{rgb}{0.7725, 0.3529, 0.0667}

\begin{document}
\title{ArtMentor: AI-Assisted Evaluation of Artworks to Explore Multimodal Large Language Models Capabilities}
\author{Chanjin Zheng}
\email{chjzheng@dep.ecnu.edu.cn}
\orcid{0000-0003-1232-0020}
\affiliation{%
  \institution{Shanghai Institute of Artificial Intelligence for Education, East China Normal University}
  \city{Shanghai}
  \country{China}
}
\affiliation{%
  \institution{Faculty of Education, East China Normal University}
  \city{Shanghai}
  \country{China}
}
\author{Zengyi Yu}
\orcid{0009-0003-4571-340X}
\affiliation{%
   \institution{Faculty of Education, East China Normal University}
  \city{Shanghai}
  \country{China}
}
\email{202105720431@zjut.edu.cn}
\affiliation{%
  \institution{College of Education, Zhejiang University of Technology}
  \city{Hangzhou}
  \state{Zhejiang}
  \country{China}
}
\author{Yilin Jiang}
\orcid{0009-0004-3179-6969}
\email{zjut_jiangyilin@163.com}
\affiliation{%
  \institution{College of Education, Zhejiang University of Technology}
  \city{Hangzhou}
  \state{Zhejiang}
  \country{China}
}

\author{Mingzi Zhang}
\orcid{0009-0006-8539-5559}
\affiliation{%
   \institution{Faculty of Education, East China Normal University}
  \city{Shanghai}
  \country{China}
}
\email{windyday@zjnu.edu.cn}
\affiliation{%
  \institution{College of Education, Zhejiang Normal University}
  \city{Jinhua}
  \state{Zhejiang}
  \country{China}
}

\author{Xunuo Lu}
\orcid{0009-0008-9174-393X}
\email{13968860822@163.com}
\affiliation{%
  \institution{School of Economy, Zhejiang University of Technology}
  \city{Hangzhou}
  \state{Zhejiang}
  \country{China}
}

\author{Jing Jin}
\orcid{0009-0006-5986-3363}
\affiliation{%
  \institution{School of Education, Zhejiang Normal University}
  \city{Jinhua}
  \state{Zhejiang}
  \country{China}
}
\email{383230730@qq.com}
\affiliation{%
  \institution{Tianchang Guanchao Primary School}
  \city{Hangzhou}
  \state{Zhejiang}
  \country{China}
}

\author{Liteng Gao}
\orcid{0009-0000-4363-2030}
\email{2335060610@st.usst.edu.cn}
\affiliation{%
  \institution{School of Artificial Intelligence Science and Technology, University of Shanghai for Science and Technology}
  \city{Shanghai}
  \country{China}
}

\thanks{$^{*}$These authors contributed equally: Chanjin Zheng, Zengyi Yu, Yilin Jiang.}  
\thanks{$^{\dag}$Corresponding author: Chanjin Zheng (Email at chjzheng@dep.ecnu.edu.cn).} 
\renewcommand{\shortauthors}{Zheng et al.}

\begin{abstract}
Can Multimodal Large Language Models (MLLMs), with capabilities in perception, recognition, understanding, and reasoning, act as independent assistants in art evaluation dialogues? Current MLLM evaluation methods, reliant on subjective human scoring or costly interviews, lack comprehensive scenario coverage. This paper proposes a process-oriented Human-Computer Interaction (HCI) space design for more accurate MLLM assessment and development. This approach aids teachers in efficient art evaluation and records interactions for MLLM capability assessment. We introduce \dataset, a comprehensive space integrating a dataset and three systems for optimized MLLM evaluation. It includes 380 sessions from five art teachers across nine critical dimensions. The modular system features \textit{entity recognition}, \textit{review generation}, and \textit{suggestion generation} agents, enabling \textit{iterative upgrades}. Machine learning and natural language processing ensure reliable evaluations. Results confirm GPT-4o’s effectiveness in assisting teachers in art evaluation dialogues. Our contributions are available at \website.
\end{abstract}



\maketitle

\begin{figure*}[t]
  \centering
  \includegraphics[width=\linewidth]{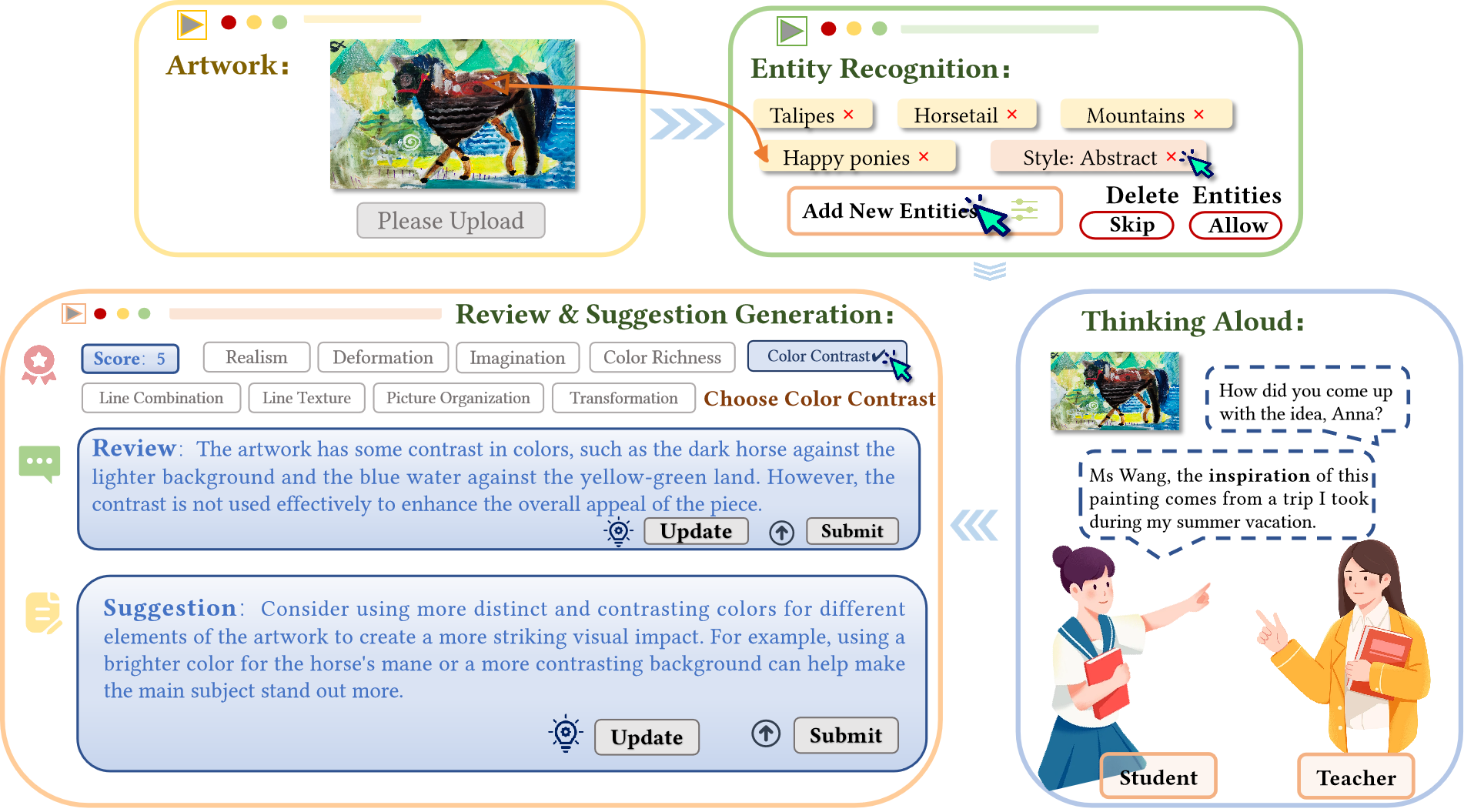}
  \caption{A multi-agent data collection system from \dataset~specifically designed to assess the GPT-4o's assistance capabilities in art evaluation. It captures interactions across 380 evaluation sessions involving five art teachers and three agents of GPT-4o. 
}
  \label{fig:1}
\end{figure*}
\section{INTRODUCTION}
Multimodal large language models (MLLMs), by seamlessly integrating various data types such as text and images, possess capabilities in multimodal perception, recognition, understanding, and reasoning \cite{huang2024survey}. 
Recent MLLMs, including GPT-4o \cite{achiam2023gpt}, Gemini \cite{team2023gemini,reid2024gemini}, and Claude 3 \cite{biswas2024robustness}, excel in tasks like image recognition, visual question answering, cross-modal retrieval, and video understanding \cite{zhang2024mm,yin2023survey}. These strengths in visual analysis and language generation offer significant potential for advancing art education assessment \cite{seo2022toward,zhao2024revolutionizing}.
In the realm of art education, particularly artwork evaluation, the integration of the assessment process with broader dialogue-based education is pivotal \cite{hubard2010three, kelaher2014evaluating}. This approach enhances students’ understanding and mitigates concerns of dehumanization. At the intersection of arts, education, and AI, a key question arises: \textit{Can MLLMs function as independent entities in collaborative evaluation processes}, effectively supporting teachers and enriching educational experiences?
In artwork evaluation, MLLMs can identify visual elements ("\textit{entities}") such as trees, faces, and art styles \cite{chen2022children}. Based on entity recognition results, MLLMs can generate artwork reviews and suggestions, functioning like an "\textit{ArtMentor}". This raises a crucial question: \textbf{How effectively can MLLMs assist elementary art teachers with \textit{entity recognition}, \textit{review generation}, and \textit{suggestion generation}?}

To assist elementary art teachers effectively, it is crucial to understand MLLMs' capabilities and limitations within specific HCI contexts \cite{cetinic2022understanding,grba2022deep}. This includes both their general performance and their behavior in educational settings \cite{qiao2022initial}. MLLMs can identify entities in artworks, link them to historical art movements, and analyze color schemes or compositions \cite{kadish2021improving,bengamra2024comprehensive}. However, their effectiveness relies heavily on the HCI methods used \cite{martinez2021developing}. In specialized HCI art environments like elementary art classrooms, the depth and accuracy with which MLLMs comprehend artistic elements are vital. 

To better tailor MLLMs for educational purposes, we have streamlined the assistance process into three distinct phases.  
Each session starts with an uploaded artwork, followed by the MLLM's automatic recognition of entities within the artwork. Incorporating "Thinking Aloud" \cite{eccles2017think} and "Protocol Analysis" \cite{ericsson2017protocol}, our system also encourages students to create audio recordings that elucidate their creative ideas and critically analyze their own work for art teachers.
Art teachers refine the entities, including art style, until accurate. After entity recognition, MLLMs review the artworks and assign scores. HCI methods verify the appropriateness of these scores, after which MLLMs provide suggestions for modifying the artwork. Each stage of this process is managed by a designated agent—namely, the \textit{entity recognition agent}, \textit{review generation agent}, and \textit{suggestion generation agent}. This modular approach supports \textit{iterative updates} and targeted enhancements, ensuring adaptability and effectiveness in educational applications. This multi-agent framework raises key questions: Can the \textit{entity recognition agent} accurately identify the themes in student artwork? How effective are the \textit{review generation agent} and \textit{suggestion generation agent} in enhancing the artistic process? Systematically addressing these questions will offer insights for refining our design. Without clear answers, assumptions about MLLMs’ effectiveness in assisting teachers remain speculative.

To address these challenges, our study draws inspiration from Mina Lee’s works \cite{lee2022coauthor, lee2024design}, particularly her focus on evaluating the writing capabilities of LLMs (GPT-3) through \textbf{HCI datasets} at CHI 2022, and her subsequent proposal of a \textbf{design space} for systematically exploring intelligent interactive writing assistants at CHI 2024. Building on this foundation, we adopt an MLLM-driven approach that collects process-oriented data during interactions between art teachers and MLLMs, expanding the scope to the multimodal domain and breaking down artistic evaluation capabilities into multiple sub-abilities.

This dynamic and ongoing data type authentically reflects the MLLM’s capabilities by capturing the evolving interaction, thus avoiding the biases of result-oriented evaluations \cite{john1996using}. In HCI research, such process-oriented data is challenging to fabricate, offering a more reliable assessment of the model’s performance in educational contexts. However, due to factors like context variability and decoding parameters \cite{li2024seed,zhang2023m3exam}, testing these capabilities poses significant challenges in HCI settings. For example, after multiple interactions between an elementary school art teacher and MLLMs, repeatedly modifying and refining reviews and suggestions on student artworks, how can we accurately identify and describe the model's specific contribution to the final reviews and suggestions? How can we quantify the extent to which the MLLMs meet the art teacher's assistance needs?

The \textit{design and analysis of HCI spaces}, encompassing datasets, multi-agent data collection systems and so on, are essential for addressing the evaluation challenges faced by MLLMs in educational assistance contexts. To gather raw data, we utilized GPT-4o, a representative MLLM, and developed the \dataset, as shown in Figure.~\ref{fig:1}. Specifically, the \dataset~space allows for an in-depth analysis of GPT-4o’s ability to assist teachers across nine key dimensions: realism, deformation, imagination, color richness, color contrast, line combination, line texture, picture organization, and transformation. Consequently, we organize both the review generation and suggestion generation agents into nine distinct 
\textit{dimensions}

\textbf{We demonstrate that the metrics designed to integrate HCI datasets with machine learning and natural language processing effectively quantify the assistance capabilities of MLLMs.} Specifcally, we adapt standard machine learning metrics such as accuracy, precision, recall, and F1-score to evaluate entity recognition capabilities, detailed in section \ref{subsec:entities rec}. Additionally, we introduce score acceptance Models outlined in section \ref{subsec:sam}. Drawing from natural language processing, we assess the generation of reviews and suggestions using two criteria: text modification length and text similarity, as discussed in section \ref{subsec:tam}. Lastly, we evaluate art style sensitivity to gauge the acceptance of art styles, which is elaborated in section \ref{subsec:ass}.

This paper makes three contributions: \textbf{(1)} We introduce a multi-agent space named \dataset, which effectively collects process-oriented HCI datasets to mitigate the fabrication issues often found in result-oriented data. The dataset and code for this space are freely accessible at \website; \textbf{(2)} We develop comprehensive evaluation metrics by integrating insights from machine learning, natural language processing, and HCI to holistically assess the assistance capabilities of MLLMs; \textbf{(3)} Through extensive data collection and analysis, We identified underperforming dimensions in the multi-agent system and proposed iterative enhancements to improve overall performance. \textit{These contributions pave the way for a more refined, process-oriented, and versatile approach to the evaluation of MLLMs' capabilities.}

\section{RELATED WORK}
\subsection{Capability Evaluation of MLLMs}
\subsubsection{Types of Capability.}
The exploration of the capabilities and limitations of MLLMs is crucial for designing effective multimodal interactions. Directly aligning with how MLLMs assist in \textit{entity recognition}, \textit{review generation} (including \textit{score}), and \textit{suggestion generation} for elementary art teachers, this foundational exploration is divided into four key sections~\cite{huang2024survey}:
\begin{itemize}[leftmargin=*]
    \item \textbf{Multimodal Perception:} Examines MLLMs' understanding of spatial and relational dynamics within data from different modalities. This includes: 
    (1) Object localization, which involves determining the position and orientation of objects within scenes, crucial for spatial awareness~\cite{chen2024gmai,yu2023mm}; 
    (2) Object relation, identifying spatial and contextual relationships between objects~\cite{liu2023mmbench,bai2023qwen}; 
    (3) Object interaction, recognizing interactions that involve actions, movements, or functional relationships within a visual context~\cite{chen2024plug,zhao2023vlchecklist}. \textit{This multifaceted approach corresponds to our system's capability to facilitate holistic assessment (art style recognition), capturing the intricate interplay of various artistic elements.}

    \item \textbf{Multimodal Recognition:} Focuses on the identification and classification of entities, actions, and attributes across different modalities, which includes: 
    (1) Concept recognition, assessing models' ability to categorize objects, actions, and scenes from varied sensory inputs~\cite{liu2023mmbench,li2023seed,yu2023mm}; 
    (2) Attribute recognition, evaluating the detection of styles, emotions, and quantities across different modalities~\cite{liu2023mmbench,awadalla2023openflamingo}; 
    (3) Action recognition, interpreting actions within various contexts~\cite{liu2023mmbench,dai2023instructblip}; 
    (4) Text recognition, determining the ability to transcribe text from images, vital for processes like automated documentation~\cite{liu2023mmbench,achiam2023gpt}. \textit{This aligns with our entity recognition process in the \dataset~space.}
    
    \item \textbf{Multi-modal Understanding:} This section evaluates MLLMs on their capability to process and make sense of data from multiple sensory inputs, extending beyond textual information, to provide a comprehensive understanding of multimodal data integration~\cite{huang2024survey}. \textit{This principle aligns with our system's capability to generate in-depth reviews and nuanced scores for individual dimensions of art evaluation, demonstrating a profound understanding of specific artistic aspects.}
    
    \item \textbf{Multimodal Reasoning:} Investigates how MLLMs infer logical conclusions from multimodal data. This section covers:
    (1) Commonsense reasoning, which evaluates models' ability to apply knowledge to interpret interactions and relationships within images~\cite{lin2024draw,yuan2024ospreypixelunderstandingvisual};
    (2) Relation reasoning, testing understanding of social, physical, or natural relations among various elements~\cite{liu2024ii,reid2024gemini};
    (3) Logic reasoning, assessing the application of logical principles in analyzing and interpreting multimodal information~\cite{liu2023mmbench,bai2023touchstone}. \textit{This is closely related to how we generate suggestions in the \dataset~space based on the assessed dimensions.}
\end{itemize}
\subsubsection{Methods to Evaluate.}
The evaluation of MLLMs encompasses several methodologies that ensure a comprehensive evaluation of their capabilities. These methods are divided into three primary categories~\cite{huang2024survey}:
\begin{itemize}[leftmargin=*]
    \item \textbf{Human Evaluation:} Human evaluators play a crucial role in assessing the capabilities of MLLMs, especially in tasks that demand high comprehension levels and are challenging to quantify using standard metrics. The evaluation focuses on multiple dimensions including: 
    (1) Relevance, assessing whether the responses align with the intended instructions~\cite{liu2023llava}; 
    (2) Coherence, determining if the responses are logically structured and consistent; 
    (3) Fluency, evaluating the naturalness and grammatical correctness of the generated outputs.

    \item \textbf{GPT-4 Evaluation:} To complement human evaluation and address its resource-intensive nature, the instruction-following capabilities of GPT-4 are used to efficiently evaluate model-generated outputs. GPT-4 assesses the MLLMs on dimensions such as helpfulness, relevance, accuracy, and detail, scoring them on a scale from 1 to 10, where higher scores indicate better performance. This approach not only provides scores but also detailed explanations for the evaluations, offering insights into the model's strengths and areas for improvement~\cite{liu2023llava, achiam2023gpt}.

    \item \textbf{Metric Evaluation:} While qualitative insights from human and GPT-4 evaluations are valuable, traditional metrics are essential for quantitatively assessing MLLM performance. These metrics provide standardized and objective measurements across various tasks:
    (1) For recognition capabilities, metrics like Accuracy and Average Precision are utilized~\cite{li2023seed, li2023llava, lau2018dataset}; 
    (2) For perception capabilities, measures such as mIoU, mAP, and Dice are adopted~\cite{dai2017scannet}; 
    (3) For evaluating text or image generation capabilities, metrics such as BLEU, ROUGE, and METEOR are widely employed~\cite{kim2019audiocaps, chen2015microsoft}, providing clear indicators of a model's performance in various applications.
\end{itemize}

While human evaluation offers insightful perspectives, it is inherently subjective and costly, with GPT-4 assessments potentially varying due to fluctuations in prompts and parameters. Furthermore, Mina Lee and colleagues have deliberated on two methodologies for investigating the generative capacities of large language models (LLMs): contextual inquiry and interaction logging analysis. Contextual inquiry, through interviews, provides profound insights albeit with limited generalizability \cite{lee2022coauthor}; interaction logging analysis, though broad in scope, lacks depth. Previous research has largely been confined to specific tasks and settings. We aim to synthesize these approaches, customizing them for multi-modal tasks using MLLMs, thereby addressing the limitations of interviews and validating the model across a wider spectrum, offering a more comprehensive evaluation and insights for future research. Consequently, \textit{we integrate traditional machine learning metrics with our innovative natural language processing techniques to deliver a nuanced, robust, and reliable assessment of MLLMs.}
\subsection{Process-oriented HCI Datasets in Education}
\subsubsection{The Growing Focus on Educational Process Mining in HCI} In the educational domain, the process of learning is often considered more critical and analytically valuable than the final outcomes \cite{vahdat2015learning}. Capturing this process, however, poses significant challenges due to the complexity of documenting and analyzing process-oriented data, which tends to be sparse and less frequently analyzed. Common applications of process mining techniques have been demonstrated in online assessment data to analyze the processes involved in answering questions or requesting student feedback \cite{pechenizkiy2009process}. Additionally, frameworks integrating educational process data mining have been introduced to facilitate the handling of interactive process data and assist educators in analyzing educational processes based on formal modeling \cite{trcka2009local}. Despite these advancements, the collection and analysis of such data remain labor-intensive and inherently complex. The complexity of process-oriented data in educational settings correlates strongly with concepts such as simplicity, ease of use, uncertainty, and the context of application, making it a focal point for HCI designers \cite{vahdat2015learning}. These challenges underscore the need for innovative approaches in HCI to enhance the usability and effectiveness of process mining tools in educational environments. \textit{Consequently, the significance and ongoing challenges of Educational Process Mining in HCI have garnered increasing attention, highlighting the urgency for developing more efficient and accessible tools.}
\subsubsection{Emerging Trends in Process-Oriented Artwork Evaluation}
In the field of artwork evaluation, process-oriented approaches are beginning to take shape. Currently, there are very few widely validated methods for automated process-oriented visual arts assessment. In fact, result-oriented methods are also scarce; one of the few examples is the Torrance Tests of Creative Thinking—Drawing Task, which scores creativity using artificial neural networks \cite{cropley2022automated}. One emerging process-oriented method involves providing an initial artwork that allows students to further develop the piece by adding patterns. The evaluation is then conducted based on both the initial and final artworks \cite{patterson2024audra}. This method attempts to document the creative process of students' artwork creation as much as possible, but recording the entire creative process remains a significant challenge. This raises an important question: \textit{Can the process of artwork evaluation under AI assistance be effectively documented?} Through the design of \dataset, we aim to facilitate interaction between educators and the system, deepening their understanding of students' artworks, thereby advancing the development of HCI in the domain of art education.

\subsection{Spaces for Iterative Multi-Agent Upgrades}
\subsubsection{Fractionalization and Dominance.}
The \textit{CoAuthor} study has significantly contributed to the HCI community by highlighting the generative capabilities of large language models  (LLMs) in creative and argumentative writing contexts \cite{lee2022coauthor}. While \textit{CoAuthor} effectively advocates for the curation and analysis of large interaction datasets to make these capabilities more transparent and accessible, it does not segment the creative process into distinct phases that could provide deeper, context-specific insights. Building on these viewpoints, we propose that applying a similar phased approach to the art evaluation process—identifying distinct stages like conception, development, and presentation—could refine our analyses even further. This approach, known as \textit{fractionalization}, involves dividing a complex process into manageable segments, with each segment handled by a dedicated agent. \textit{Dominance} in this context refers to the strategic control and optimization of each segment by its respective agent, ensuring that the overall system maintains coherence and maximizes efficiency. By implementing a system based on these principles, not only is the granularity of the obtained insights improved, but the overall effectiveness and adaptability of the system within dynamic HCI environments are also enhanced. Similarly, when applied to the \textit{Evaluation of Artworks}, this structured approach allows specialized agents to precisely manage different stages of artistic creation and interpretation, thereby enhancing the precision and depth of art evaluations. Consequently, our \dataset~is structured around multi-agent concepts based on fractionalization and dominance principles.
\subsubsection{Living Artifact and Iterative Upgrades.}
Following the \textit{CoAuthor} research, Lee proposed a dynamic and adaptive framework aimed at continuously enhancing technologies for writing assistance \cite{lee2024design}. This design space, developed through collaborations with experts from disciplines including HCI, Natural Language Processing, Information Systems, and Education, encompasses five key dimensions—\textit{task, user, technology, interaction, and ecosystem}—and involved a comprehensive analysis of 115 papers to map the landscape of writing assistants. The framework is designed as a \textit{living artifact}, intended to evolve through community contributions of new research, annotations, and discussions, keeping pace with advancements in the field. However, while this framework is insightful, it still relies on a traditional data collection system. Inspired by Talebirad's approach to enhancing LLMs through multi-agent systems \cite{talebirad2023multi}, our \dataset~employs a similar structure to address these limitations. By adopting a multi-agent architecture, we facilitate \textit{iterative upgrades}, enabling both the dataset and the supporting system to remain dynamic and responsive to emerging needs and developments. This strategy not only enhances the adaptability of our system but also improves its capability to handle complex tasks efficiently, reflecting the collaborative environment and knowledge exchange among intelligent agents envisioned by Talebirad.
\section{DESIGN PRINCIPLES FOR \dataset}\label{dpf}
In this section, we discuss the design of four main components of \dataset: \textit{a multi-agent data collection system, an HCI dataset, a data analysis system}, and \textit{an iterative upgrades system}. 
The design of \dataset~adheres to the principles of adjusting evaluation granularity, providing immediate feedback, and progressively approaching target capabilities \cite{anderson1995cognitive}. To design appropriate evaluation granularity, we draw inspiration from Henri Bergson’s exploration of consciousness \cite{bergson1911essai}, shifting the focus from evaluating the physical aspects of art pieces to assessing the creative abilities of the students who create them, thereby formulating nine dimensions. Immediate feedback is linked to our approach of HCI, where MLLMs continuously engage with art teachers and students, and this feedback is meticulously recorded. Instead of directly evaluating the art pieces, we divide the process into multiple sub-processes, including entity recognition, commentary generation, and suggestion generation, thereby embodying the principle of progressively approaching target skills.
\subsection{Adjusting Evaluation Granularity}
Our research extends prior studies by designing a multi-agent data collection system, which serves as a core component of \dataset ~\cite{song2007multi}. Art, as a unique form of expression, is deeply rooted in the pursuit of the essence of life, providing individuals with spiritual fulfillment \cite{Zhang2020}. In the era of big data, the evaluation of artworks faces several technical challenges, primarily manifested in four areas: data dependency, limitations on creativity, emotional reduction, and the loss of intrinsic meaning \cite{Jin2024}. For instance, instrumental rationality often causes data to become a rigid constraint on both teachers' and students' creative interpretations of art.

To address these challenges, our system is based on the principle of relational self-expression, which elevates the evaluation process beyond mere objectivity. The core concept of relational self-expression draws inspiration from Henri Bergson's philosophy of creative evolution, emphasizing the notion that relation precedes individuality. To fully implement this idea, we incorporate insights from Bergson's exploration of consciousness \cite{bergson1911essai}. Bergson's emphasis on direct experience and pre-reflective consciousness seeks to capture the undistorted reality of conscious states. Similarly, our system aims to evaluate artworks in their most authentic form. The system integrates multiple data types, focusing on images, audio, and textual data. Moreover, the evaluation process is implemented within a CHI  framework, leveraging GPT-4o to assist art teachers in scoring. This hybrid approach combines machine intelligence with human expertise, ensuring both precision and contextual depth in artistic evaluation.

Our evaluation framework encompasses multiple dimensions, each designed to capture the unique characteristics of artworks. These dimensions include formative creativity (\textbf{\textcolor{art-realism}{realism}}, \textbf{\textcolor{art-deformation}{deformation}}, \textbf{\textcolor{art-imagination}{imagination}}), color expressiveness (\textbf{\textcolor{art-color-richness}{color richness}}, \textbf{\textcolor{art-color-contrast}{color contrast}}), line work richness (\textbf{\textcolor{art-line-combination}{line combination}}, \textbf{\textcolor{art-line-texture}{line texture}}), and conceptual thinking (\textbf{\textcolor{art-picture-organization}{picture organization}}, \textbf{\textcolor{art-transformation}{transformation}}). This dimensional framework was developed by a member of our research team and represents an innovative contribution to the field, though the corresponding work is yet to be formally published.

\begin{itemize}[leftmargin=*]
    \item \textbf{\textcolor{art-realism}{Realism}}: Evaluates the artwork's ability to realistically reproduce subjects, capturing the precision and accuracy of representation \cite{biswas2021realism}.
    \item \textbf{\textcolor{art-deformation}{Deformation}}: Assesses the artwork's capacity to transform and recreate reality, reflecting artistic innovation and reinterpretation \cite{sfarra2014discovering}.
    \item \textbf{\textcolor{art-imagination}{Imagination}}: Examines the creativity and originality within the artwork, highlighting the artist's ability to introduce new and novel perspectives \cite{searle2016capturing}.
    \item \textbf{\textcolor{art-color-richness}{Color Richness}}: Evaluates the diversity and aesthetic harmony of the color palette used in the artwork \cite{lu2015discovering,pylypchuk2021developing}.
    \item \textbf{\textcolor{art-color-contrast}{Color Contrast}}: Assesses the visual impact and vibrancy of colors, focusing on how contrasting hues interact within the composition \cite{zhang2021comprehensive}.
    \item \textbf{\textcolor{art-line-combination}{Line Combination}}: Evaluates the arrangement and structural coherence of lines, examining how line elements contribute to the overall form \cite{locher1999empirical}.
    \item \textbf{\textcolor{art-line-texture}{Line Texture}}: Examines the expressiveness and tactile quality of line work, exploring the textural effects achieved through line variations \cite{ding2020image}.
    \item \textbf{\textcolor{art-picture-organization}{Picture Organization}}: Evaluates the overall layout and arrangement of elements within the artwork, assessing compositional balance and spatial logic \cite{locher1999empirical}.
    \item \textbf{\textcolor{art-transformation}{Transformation}}: Assesses the artist's ability to transform abstract concepts into tangible artistic expressions, exploring the depth of conceptual execution \cite{du2020research}.
\end{itemize}

This multidimensional evaluation framework, grounded in the principle of relational self-expression, represents an AI-driven approach to artistic evaluation. It is based on the processes of data collection, organization, and accumulation, transitioning from object-centric, observable, and dynamic assessments to human-centered evaluations. By constructing a human-machine collaborative evaluation system, the framework ultimately aims to transcend the surface self and reach the deep self, fostering the creative and emotional expression of artistic life.

\subsection{Providing Immediate Feedback}
The design principles of the HCI dataset should prioritize both the evaluation outcomes and the comprehensive recording of all user-system interactions. The GOMS (Goals, Operators, Methods, and Selection Rules) model provides a systematic framework for tracking interactions, emphasizing the centrality of operators (user actions) and methods (task execution). This decomposition facilitates researchers to isolate and analyze each component, clarifying how users, such as art educators, navigate the system, refine their evaluations, and adjust feedback based on AI-assisted recommendations. \cite{lamb2017pass}.

Leveraging the GOMS  model, which emphasizes breaking down complex tasks into smaller, manageable components \cite{rowe2008rule,shi1992impact},  we advocate dividing interaction processes into distinct phases to capture decision-making and feedback loops holistically. For example, in the initial phase of artwork evaluation, the dataset must capture every interaction between the educator and the \textit{entity recognition agent}, including the system’s identification results, the educator’s revisions, and the final outcome. Subsequently, the \textit{review generation agent} records initial user ratings and comments, followed by agent revisions, user modifications, and final adjustments, culminating in the user’s submission. Finally, the \textit{suggestion generation agent} data encapsulates the suggestions provided by MLLMs for improving the artwork, which can be analyzed to understand how AI-assisted feedback is utilized and acted upon in real-world educational contexts.

Each phase is systematically recorded, ensuring a comprehensive dataset for analyzing and improving MLLM capabilities. This comprehensive methodology is in line with the GOMS model's emphasis on evaluating human-system interaction by understanding not only the results, but also the cognitive steps taken throughout the process \cite{janssen2015strategic}. Moreover, the ability to refine each agent’s performance based on the interaction data ensures that the dataset remains an integral tool in the ongoing evolution of MLLM-assisted artwork evaluation \cite{nemlekar2021two}. 

\subsection{Approaching Target Capabilities}
The data analysis system should enable designers to extract meaning from interactions and analyze them based on their own design goals \cite{blandford1995using}. Instead of solely depending on subjective scoring of the procedural data to evaluate MLLMs’ assistance capabilities—often prone to bias—the system facilitates a more objective, iterative evaluation process across multiple rounds. Art teachers are empowered to modify the data, prompting MLLMs to regenerate scores, reviews, and suggestions in subsequent rounds based on these adjustments. In terms of assistance capabilities, we divide them into four categories: \textit{entity recognition}, \textit{style evaluation}, \textit{scoring}, and \textit{text generation} (\textit{reviews} and \textit{suggestions}). These capabilities are detailed as follows:
\begin{itemize}[leftmargin=*]
    \item \textbf{Entity Recognition Capability}: Inspired by machine learning classification metrics, we define accuracy, recall, and other metrics based on the interaction between the entity recognition agent and art educators. Following each round of teacher-driven entity modifications (deleting or adding), we quantify the MLLMs' entity recognition capability, offering high interpretability through successive rounds of human-AI interaction.
    
    \item \textbf{Style Evaluation Capability}: We assess the MLLMs’ style evaluation capability by measuring the extent to which art educators accept or reject the styles identified by the MLLMs, calculated by whether the teacher deletes the recognized styles.
    
    \item \textbf{Scoring Capability}: We extract the scores from the reviews to evaluate MLLMs’ initial scoring of artworks, the manual scores given by teachers, and the scores regenerated by the MLLMs after human modifications. By comparing the initial MLLMs scores with teachers' scores and tracking the similarity across multiple score generation rounds, we assess the progression of MLLMs' scoring capability.

    \item \textbf{Text Generation Capability}: Both reviews and suggestions are treated as text outputs that can be evaluated using the same metrics. Utilizing natural language processing techniques, we divide the MLLMs’ text generation capability into two main parts: the degree of modification and the similarity of the text. These metrics are derived from the modification length between initial and revised texts, and the word similarity measured through tokenization and semantic analysis over successive rounds.
\end{itemize}

\begin{figure*}[t]
  \centering
  \includegraphics[width=\linewidth]{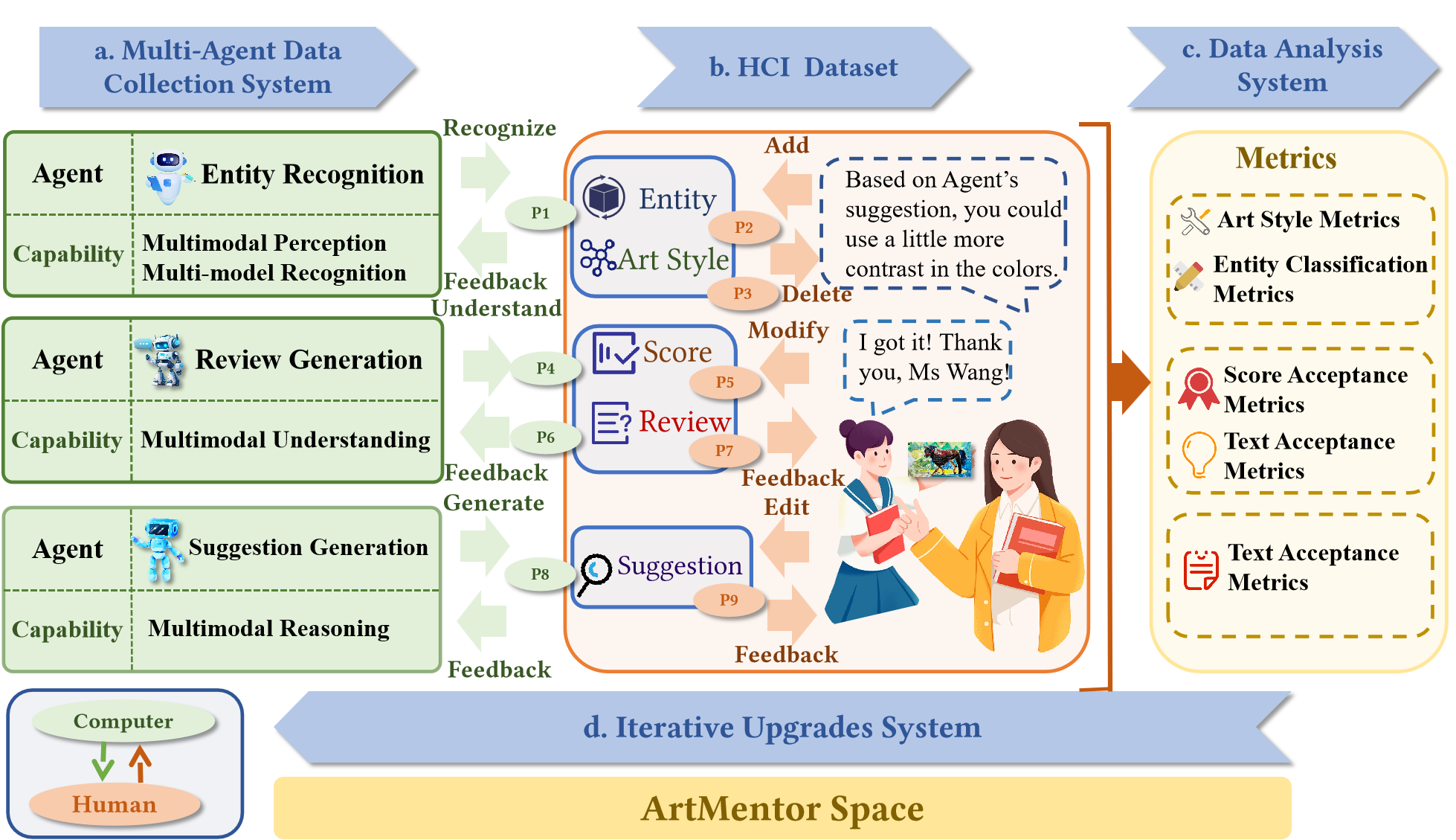}
  \caption{\dataset~Space comprises four primary components: \textbf{a. Multi-Agent Data Collection System}, \textbf{b. HCI Dataset}, \textbf{c. Data Analysis System}, \textbf{d. Iterative Upgrades System}. The Multi-Agent Data Collection System includes three agents: \textbf{E}ntity Recognition \textbf{Agent} (\textbf{E-Agent}), \textbf{R}eview Generation \textbf{Agent} (\textbf{R-Agent}), and \textbf{S}uggestion Generation \textbf{Agent} (\textbf{S-Agent}). Both R-Agent and S-Agent perform nine dimensions, such as Realism and Deformation. Additionally, we have outlined nine HCI processes (from P1 to P9), where processes initiated by the \textcolor{computer-green}{\textbf{computer}} are marked in green and those initiated by the \textcolor{human-orange}{\textbf{human}} are marked in orange. After data collection by the Multi-Agent system, we obtain an HCI dataset. We then apply five metrics to evaluate these four capabilities. Based on the evaluation results, we aim to iteratively upgrade capabilities that underperform in the future.}
  \label{fig:2}
\end{figure*}

\section{METHODOLOGY}
According to principles in Section \ref{dpf}, we develop \dataset, which includes 9 HCI processes, as shown in Figure.~\ref{fig:2}. First, the HCI dataset comprises these nine HCI processes, all of which are documented by the multi-agent data collection system and detailed in Section \ref{RPF}. Subsequently, we extract the significance of the processes into metrics, and provide feedback for iterative upgrades system through a data analysis system in Section \ref{ECF}.

\subsection{Documenting Processes of the Multi-agent Data Collection System for an HCI Dataset}\label{RPF}
\subsubsection{Art Teachers and E-Agent Interaction Processes.} In the multi-agent data collection system, the E-Agent is responsible for extracting and classifying entities from artworks using a MLLM. We demonstrate the details of Interaction Processes. 

\textcolor{computer-green}{\textit{\textbf{Process 1 (P1): Entity List Recognition.}}} The E-Agent initiates the process of recognizing entities from the given artworks. This process can be described by the following equation:
\begin{equation}
  E_{i} = Ag_{e}(ML, Art_{i}, Ent\_prm, \Theta, \Gamma),
  \label{eq:entity_extraction}
\end{equation}
where \( E_{i} \) is the set of entities recognized from the \( i \)-th artwork, \( Ag_{e} \) represents the E-Agent driven by the specific MLLM (ML), \( Art_{i} \) is the \( i \)-th artwork being analyzed, \( Ent\_prm \) is the prompt provided to the MLLM to guide the entity extraction, \( \Theta \) is a set of parameters used in the extraction process, and \( \Gamma \) is the set of dimension-related hyperparameters.  

\textcolor{human-orange}{\textit{\textbf{Process 2 (P2): Entity Right List Addition.}}} During this process, the teacher adds new correct entities that are not already present in the set \( E_{i} \). The set of entities newly added by the teacher is defined by the following equation:
\begin{equation}
  R_{i} = \{ e_{im} \mid m = 1, 2, 3, \ldots, M_{i}, \forall e_{im} \notin E_{i} \},
  \label{eq:right_entity_addition}
\end{equation}
where \( R_{i} \) represents the set of entities added by the teacher, and \( e_{im} \notin E_{i} \) for all \( m \). 

\textcolor{human-orange}{\textit{\textbf{Process 3 (P3): Entity Wrong List Deletion.}}} 
The teacher removes incorrect entities from \( E_{i} \), defined as:
\begin{equation}
  W_{i} = \{ e_{iq} \mid q = 1, 2, 3, \ldots, Q_{i}, \forall e_{iq} \in E_{i} \},
  \label{eq:wrong_entity_deletion}
\end{equation}
where \( W_{i} \) is the set of entities identified as incorrect by the teacher, and \( e_{iq} \in E_{i} \) for all \( q \).

\textit{\textbf{Final Entity List Update.}} Following the revisions of correct and incorrect entities, the final set of entities for each artwork is updated by incorporating the revised correct entities and excluding the incorrect ones. As defined by Eqs. \ref{eq:entity_extraction}, \ref{eq:right_entity_addition}, and \ref{eq:wrong_entity_deletion}, the final entity list is expressed as:
\begin{equation}
  \hat{E_{i}} = \overline{E_{i}} = E_{i} \cup R_{i} \setminus W_{i},
  \label{eq:final_entity_list_update}
\end{equation}
where \(\hat{E_{i}}\) denotes the final revised list of recognized entities. The new \( \overline{E_{i}} \) represents the updated set of entities for the \( i \)-th artwork, revised by the art teacher. After revision, \(\hat{E_{i}}\) is equivalent to \( \overline{E_{i}} \).

\subsubsection{Art Teachers and R-Agent Interaction Processes.} The R-Agent is responsible for generating reviews for artworks, which involves both scoring (ranging from 1 to 5 as integers) and textual review generation. The primary processes involved are: \textit{review generation}, \textit{review modification}, \textit{score extraction}, and \textit{score adjustment}.

\phantomsection
\textcolor{computer-green}{\textit{\textbf{Process 4 (P4): Review Generation.}}}\label{process4} The R-Agent generates a review for the \( k \)-th dimension of the \( i \)-th artwork as defined by the following equation:
\begin{equation}\label{eq:review generation}
Rev_{ik} = Ag_{r}(ML, Art_{i}, Rev\_Pmt_{k}, \hat{E_{i}}, \theta_{k}, \gamma_{k}, \overline{Rev}_{ik}, f),
\end{equation}
where \( Rev_{ik} \) is the review generated for the \( k \)-th dimension of the \( i \)-th artwork. If \( \overline{Rev}_{ik} \) is empty, the review is solely generated by the computer; if \( \overline{Rev}_{ik} \) contains content, it indicates collaboration between the human and the computer in the review process. \( Ag_{r} \) is the Review-Agent, \( ML \) denotes the specific MLLM, \( Art_{i} \) refers to the artwork in question, \( Rev\_Pmt_{k} \) is the prompt for the review, \( \hat{E_{i}} \) is the final list of entities updated through interaction with the E-Agent and humans as per Eq. \ref{eq:final_entity_list_update}, \( \theta_{k} \) and \( \gamma_{k} \) are the parameters and dimension-related hyperparameters used in generating the review, respectively, and \( f \) is the function used for extracting scores from the reviews.

\phantomsection
\textcolor{human-orange}{\textit{\textbf{Process 5 (P5): Review Modification.}}}\label{process5} During this process, the teacher modifies the generated review. The modifications are quantified by the following equation:
\begin{equation}\label{eq:review modification}
\overline{Rev}_{ik} = Rev_{ik} \cup Ins_{ik} \setminus Del_{ik},
\end{equation}
where \( \overline{Rev}_{ik} \) represents the review following human modifications. This review may be co-generated by both the computer and the human if \( Rev_{ik} \) as defined in Eq. \ref{eq:review generation} is not empty. If \( Rev_{ik} \) is empty, it indicates no computer involvement in the initial review generation. \( Ins_{ik} \) denotes the review content inserted by the teacher for the \( k \)-th dimension of the \( i \)-th artwork, and \( Del_{ik} \) denotes the review content removed from the same dimension.

\textcolor{computer-green}{\textit{\textbf{Process 6 (P6): Score Extraction.}}}
Scores are extracted from the generated review as follows:
\begin{equation}\label{eq:score extraction}
  S_{ik} = f(Rev_{ik}),
\end{equation}
where \( S_{ik} \) represents the score for the \( i \)-th artwork along the \( k \)-th dimension. The function \( f \) extracts the score from the review \( Rev_{ik} \), specifically deriving from the procedures outlined in Process 4.

\textcolor{human-orange}{\textit{\textbf{Process 7 (P7): Score Adjustment.}}}
This process involves the modification of scores as detailed below:
\begin{equation}\label{eq:score adjustment}
  \overline{S}_{ik} = S_{ik} - \Delta s,
\end{equation}
where \( \overline{S}_{ik} \) denotes the adjusted score for the \( i \)-th artwork on the \( k \)-th dimension, and \( \Delta s \) represents the human modification, which may be either an increase or a decrease. All scores are integers within a five-point scale.

\textit{\textbf{Final Review Submission.}} In the concluding process, scores and reviews are finalized through collaborative efforts between human and computer. The initial reviews and scores produced by the computer undergo thorough scrutiny and adjustments by the teacher, enhancing their accuracy and reliability. The final review is denoted as \( \hat{Rev}_{ik} \) and includes the final score \( \hat{S}_{ik} \).

\subsubsection{Art Teachers and S-Agent Interaction}
The S-Agent generates and revises artwork suggestions, involving processes like \textit{Suggestion Generation}, \textit{Modification}, and \textit{Final Submission}. Both reviews and suggestions are text-based with similar interaction processes.

\textcolor{computer-green}{\textit{\textbf{Process 8 (P8): Generation}}}:
\begin{equation}\label{eq:generation}
Sug_{ik} = Ag_{s}(ML, Art_{i}, \ldots, \overline{Sug}_{ik}),
\end{equation}
where \(Sug_{ik}\) is the suggestion, \(Ag_{s}\) is the S-Agent, and \(\overline{Sug}_{ik}\) is the co-generated suggestion.

\textcolor{computer-green}{\textit{\textbf{Process 9 (P9): Modification}}}:
\begin{equation}\label{eq:modification}
\overline{Sug}_{ik} = Sug_{ik} \cup InsSug_{ik} \setminus DelSug_{ik},
\end{equation}
where \(\overline{Sug}_{ik}\) is the final suggestion, \(InsSug_{ik}\) is the inserted content, and \(DelSug_{ik}\) is the removed content.

\textit{\textbf{Final Submission}}: The finalized suggestions \( \hat{Sug}_{ik} \), crafted through collaboration, are submitted.

\subsection{Evaluation from the Data Analysis System for Iterative Upgrades System}\label{ECF}
To evaluate assistance capabilities of MLLMs, our data analysis system covers diverse metrics: \textit{Entity Classification Metrics} (inspired by machine learning classification metrics), \textit{Art Style Metrics}, \textit{Score Acceptance Metrics}, and \textit{Text Acceptance Metrics}. The insights gained from these metrics guide the iterative enhancement of capabilities.
\subsubsection{Entity Classification Metrics for E-Agent}\label{subsec:entities rec}
For the E-Agent, we demonstrate the design of entity classification metrics derived from the confusion matrix. We define \textit{True Positive (TP)}, \textit{Misrepresentation (MR)}, \textit{False Positive (FP)}, and \textit{False Negative (FN)} as follows:
\begin{itemize}[leftmargin=*]
    \item \textbf{TP}: The number of entities correctly identified in the artwork by the E-Agent reflecting accurately recognized entities. The calculation is expressed as:
    \begin{equation}
    TP = |E_i| - |W_i|.
    \end{equation}

    \item \textbf{MR}: The number of entities incorrectly identified as other entities by the E-Agent, e.g., a "horse" misidentified as a "donkey". The calculation follows:
    \begin{equation}\label{MR}
    MR = \min(|W_i|, |R_i|).
    \end{equation}

    \item \textbf{FP}: The number of non-existent entities recognized by the E-Agent, e.g., mistakenly identifying nonexistent water in an artwork. The calculation is adjusted for misrepresentations:
    \begin{equation}
    FP = \max(|W_i| - MR, 0)\label{eq:fp}. 
    \end{equation}

    \item \textbf{FN}: The number of entities overlooked by the E-Agent. This involves entities present in the artwork but not recognized, calculated by:
    \begin{equation}
    FN = \max(|R_i| - MR, 0)\label{eq:fn}. 
    \end{equation}
\end{itemize}

Based on these definitions, we redefine the metrics of \textit{accuracy}, \textit{precision}, \textit{recall}, and \textit{F1-score} to evaluate the E-Agent's entity recognition capabilities:
\begin{equation}
\begin{aligned}  
Accuracy &= \frac{TP}{TP + FP + FN + MR},\\
Precision &= \frac{TP}{TP + FP + MR}, \\
Recall &= \frac{TP}{TP + FN + MR},\\
F1 &= \frac{2 \times Precision \times Recall}{Precision + Recall}.
\end{aligned}
\label{eq:precision}
\end{equation}

These calculations provide a comprehensive evaluation of MLLMs' performance in entity recognition within artworks, facilitating targeted improvements to the model.
\subsubsection{Art Style Metrics for E-Agent}\label{subsec:ass}
We conceptualize art style as a distinct entity and introduce the \textbf{Art Style Sensitivity (ASS)} metric to assess the E-Agent's capability to accurately recognize and evaluate various art styles. The metric is defined as follows:
\begin{equation}
\text{ASS} = 1 - \frac{D}{N},
\end{equation}
where \( N \) represents the total number of art styles identified, and \( D \) denotes the number of incorrect recognitions as flagged and corrected by the art teacher.
\subsubsection{Score Acceptance Metrics for R-Agent}\label{subsec:sam}
As the scoring of artworks becomes increasingly prevalent in automated assessment systems, it is crucial to evaluate the scoring capabilities of the R-Agent. We focus on three key aspects: First: the \textit{score difference} between \textbf{initial} scores provided by MLLMs and the scores assigned by art teachers in subsequent rounds (\textcolor{computer-green}{\hyperref[process4]{\textit{\textbf{Process 4}}}} to \textcolor{human-orange}{\hyperref[process5]{\textit{\textbf{Process 5}}}}). Second, the \textit{consistency of scoring} between MLLMs and art teachers across all rounds. Third, the \textit{volatility of scores} either from art teachers or MLLMs.

\textbf{Score Difference (SD)}: This metric quantifies the deviation between the initial MLLM-generated scores and the modified scores assigned by art teachers in subsequent rounds (\textcolor{human-orange}{\hyperref[process5]{\textit{\textbf{Process 5}}}}). The equation is defined as:
\begin{equation}
\text{SD} = \frac{1}{N} \sum_{i=1}^{N} \left| S_{\text{MLLM}}^{(1)} - S_{\text{art teacher}}^{(i)} \right|,
\end{equation}
where \( S_{\text{MLLM}}^{(1)} \) represents the initial MLLM score, \( S_{\text{art teacher}}^{(i)} \) denotes the art teacher’s modified score in the \( i \)-th round, and \( N \) is the total number of rounds.

\textbf{Score Consistency (SC)}: This metric evaluates the alignment between art teachers' scores and those of MLLMs by measuring the Spearman correlation coefficient (\(\rho\)) between the scores given by MLLMs and the scores modified by art teachers. The Spearman coefficient is a non-parametric measure of rank correlation that assesses the monotonic relationship between two variables. It is calculated as follows:
\begin{equation}
\text{SC} = \rho(S_{\text{MLLM}}, S_{\text{art teacher}}),
\end{equation}
where \( S_{\text{MLLM}} \) and \( S_{\text{art teacher}} \) are vectors representing the series of scores from MLLMs and the corresponding modified scores by art teachers, across all rounds.

\textbf{Score Volatility (SV)}: This metric quantifies the stability of scoring by either art teachers or MLLMs across all rounds, reflecting the consistency of the scoring process. It is calculated using the standard deviation:
\begin{equation}
\text{SV} = \text{std}(S_{\text{scorer}}),
\end{equation}
where \( S_{\text{scorer}} \) represents the vector of scores modified by a specific scorer, either an art teacher or an MLLM. This metric is crucial for identifying instances of erratic scoring behavior that could undermine the reliability of the analysis.
\subsubsection{Text Acceptance Metrics for R-Agent and S-Agent}\label{subsec:tam}
For R-Agent and S-Agent, natural language generation represents a core capability, producing both reviews and suggestions in textual form. However, analyzing text within HCI contexts can be challenging. To address this, we propose evaluating text through two key metrics: text length and word similarity. These metrics respectively provide insights into the conciseness and relevance of the generated text.

\textbf{Text Modification Rate (TMR)}: This metric quantifies the acceptance of MLLM-generated text (e.g., reviews, suggestions) by calculating the ratio of characters modified by the art teacher to the original text. The formula is defined as:
\begin{equation}
\text{TMR} = \frac{1}{N} \sum_{i=1}^{N} \frac{\text{len}(T_{\text{MLLM}}^{(i)}) - \text{len}(T_{\text{removed}}^{(i)})}{\text{len}(T_{\text{MLLM}}^{(i)})+\text{len}(T_{\text{added}}^{(i)})},
\end{equation}
where \( T_{\text{added}}^{(i)} \) and \( T_{\text{removed}}^{(i)} \) represent the number of characters added and removed by the art teacher in the \( i \)-th round, respectively, and \( T_{\text{MLLM}}^{(i)} \) is the original number of characters in the MLLM-generated text.

\textbf{Text Similarity (TS)}: We assess the similarity between the original MLLM-generated text and the modified text using cosine similarity metrics. Prior to similarity computation, both texts are vectorized using the Bag of Words (BoW) model, referred to as "Wordbag" in our context. This model represents each text as a vector, where each dimension corresponds to a word in the vocabulary. The value in each dimension represents the frequency of the corresponding word in the text. The vectorization process is detailed as follows:

\begin{equation}
\begin{aligned}
T_{\text{MLLM}} = \text{Wordbag}(T_{\text{MLLM}}),\\
T_{\text{art teacher}} = \text{Wordbag}(T_{\text{art teacher}}),
\end{aligned}
\end{equation}
where \( T_{\text{MLLM}} \) and \( T_{\text{art teacher}} \) are the vectorized representations of the MLLM-generated and art teacher-modified texts, respectively.

Cosine similarity is then used to compute the degree of similarity:
\begin{equation}
\text{TS} = \text{Cosine}(T_{\text{MLLM}}, T_{\text{art teacher}}) = \frac{T_{\text{MLLM}} \cdot T_{\text{art teacher}}}{\|T_{\text{MLLM}}\| \|T_{\text{art teacher}}\|}.
\end{equation}
\section{GUIDE FOR \dataset~SPACE}
\subsection{Data Collection Interface}
The user interface and interaction design of the \dataset~space are specifically developed to provide art teachers with an intuitive and structured tool for evaluating artworks. The primary objective of this interface is to support detailed art analysis through a series of structured interaction processes.
\begin{figure}[htbp]
  \centering
  \begin{minipage}[b]{0.50\textwidth}
    \centering
    \includegraphics[width=\linewidth]{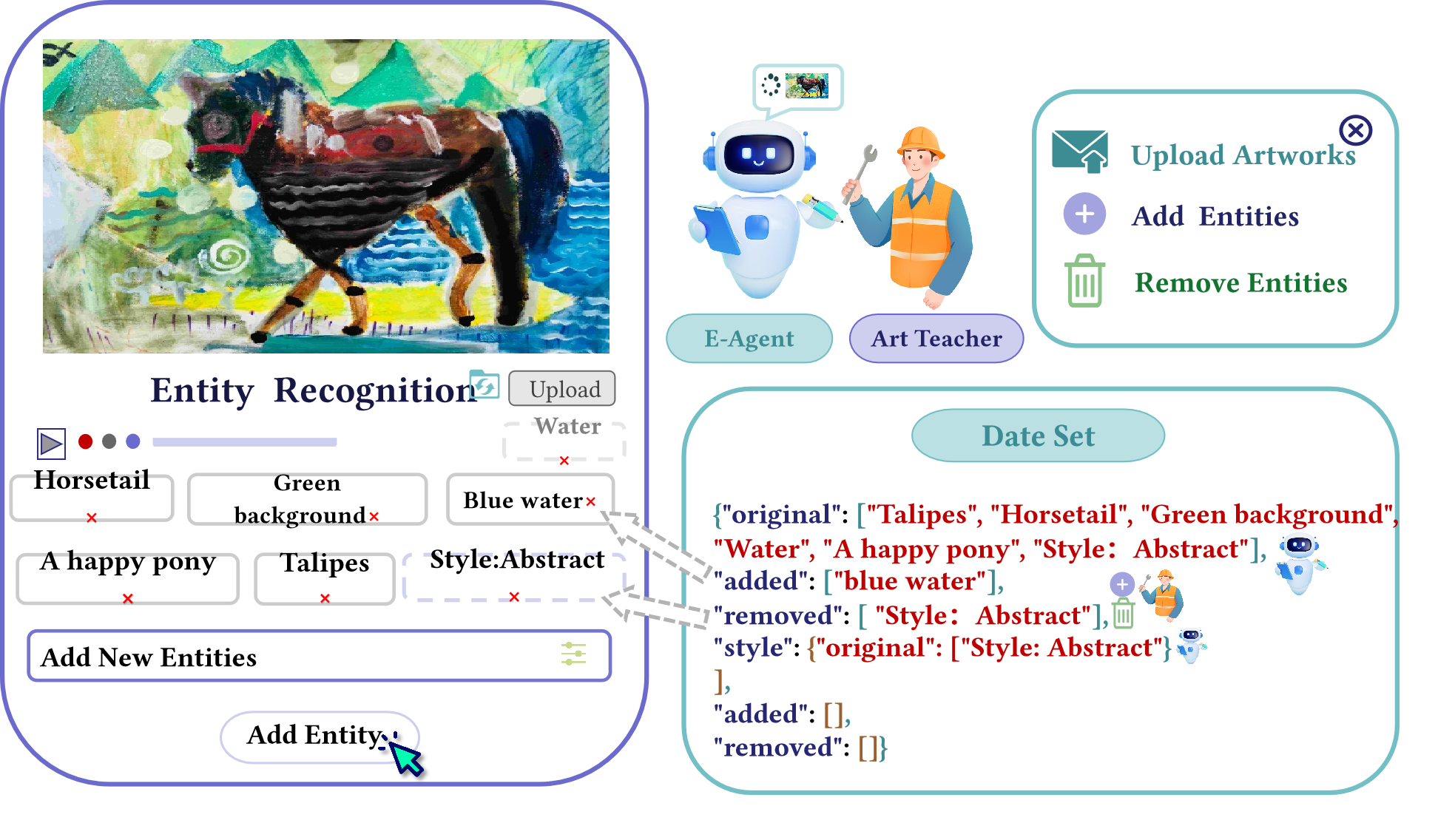}
    \caption{E-Agent and art teacher interaction collection.}
    \label{fig:interface1}
  \end{minipage}
\end{figure}

As shown in Figure. \ref{fig:interface1}, the interface allows art teachers to upload an artwork for analysis and provides multiple evaluation sections based on different artistic dimensions. After uploading the artwork, the interface displays the artwork and automatically recognizes and lists entities within the artwork. Art teachers can add or remove entities to optimize the recognition results. Finally, the HCI processes saved in JSON files.

\begin{figure}[htbp]
  \centering
  \begin{minipage}[b]{0.50\textwidth}
    \centering
    \includegraphics[width=\linewidth]{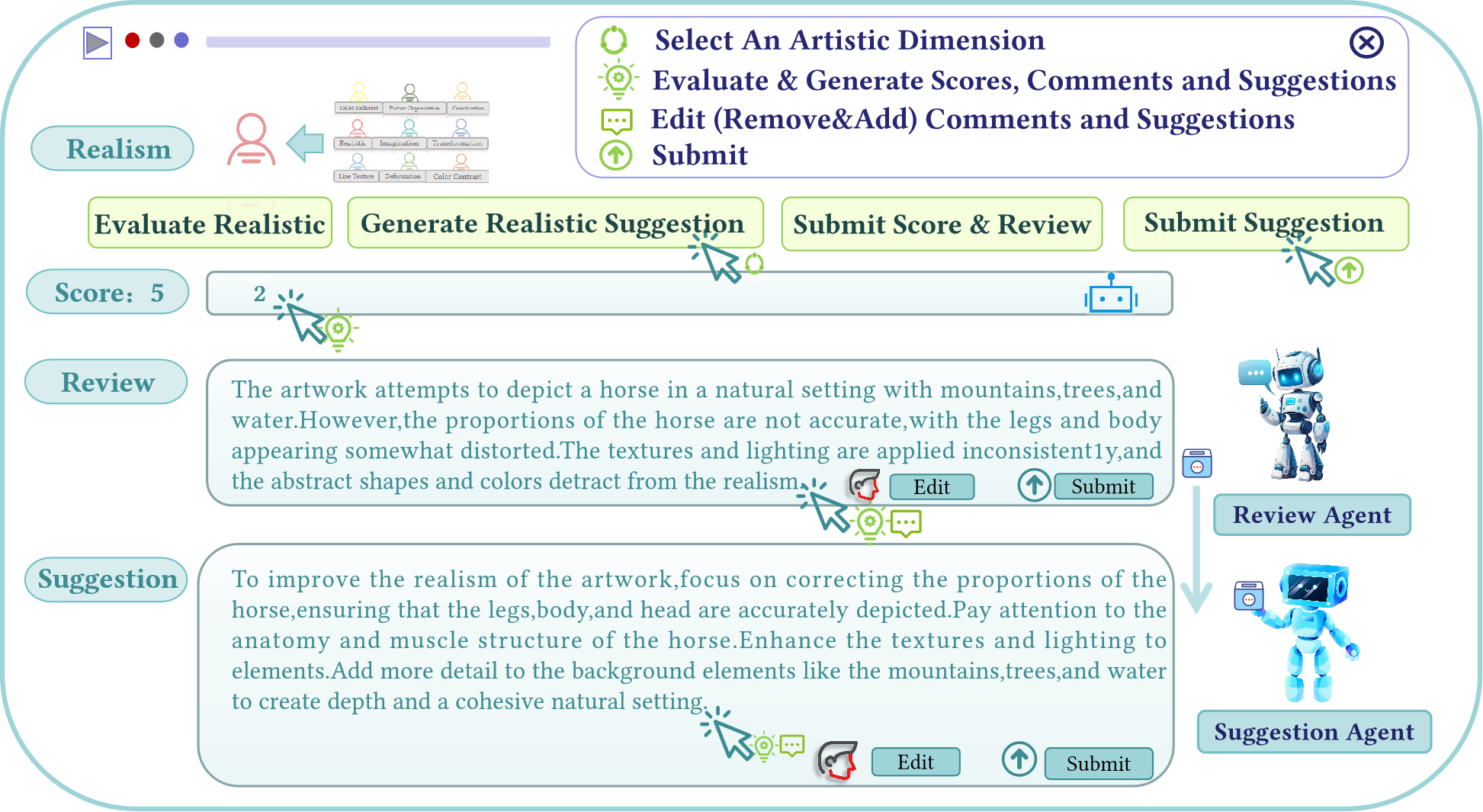}
    \caption{R \& S-Agent and art teacher interaction collection.}
    \label{fig:interface2}
  \end{minipage}
\end{figure}

As shown in Figure. \ref{fig:interface2}, the interface offers a set of categorized evaluation tools. Art teachers can select different dimensions (e.g., realism, deformation) to generate or manually input scores and reviews. Each dimension has specific evaluation tools, including buttons for generating review and suggestions from MLLMs. 
\subsection{Data Collection Procedure}
\subsubsection{Participants and Artworks}
In the \dataset~space, we recruited five art teachers from diverse educational backgrounds, including both pre-service and in-service art teachers. The selection of participants was aimed at offering a range of perspectives to facilitate a comprehensive evaluation of elementary students' artworks. The artworks used in this study were categorized into three groups: narrative illustrations (1-3), Chinese ink paintings (4-7), and artworks following the Egyptian frontal law (8-20), with each number range corresponding to its respective category.
\subsubsection{Art Evaluation Details}
Participants were provided with detailed instructions on how to interact with the \dataset~space. They were required to evaluate each piece of artwork across multiple dimensions, spending at least ten minutes on each evaluation. To enhance the multidimensional nature of the evaluation process and capture more subtle aspects of artistic expression, students were allowed to provide audio explanations of their artworks. These audio explanations allowed art teachers to gain insights into the students' creative processes, intentions, and emotional connections to their work, aligning with our principle of evaluation beyond dehumanization. The guidelines emphasized three primary requirements: (1) ensuring that each evaluation is thorough, covering multiple artistic dimensions; (2) actively using the MLLM-generated reviews and suggestions provided by the system to enhance their evaluations; (3) carefully considering the students' audio explanations to capture nuanced information about the artistic process and intent.

\subsubsection{Post-Evaluation Interviews}
After completing each evaluation session, participants were asked to participate in an interview designed to gather feedback on their experience with the \dataset~space. The interview focused on five key areas: participants' perceptions of the accuracy of the MLLM's entity recognition, the usability and user interaction experience with the space, the effectiveness of MLLM in assisting with scoring and review or suggestion generation, the MLLM's performance across different functional dimensions, and the relevance of the MLLM's suggestions in supporting elementary students' artistic development. The detailed interview questionnaire is provided in Appendix \ref{ss}. 
The content of the interviews will be used to support the result analysis, and some materials can be accessed at \href{https://github.com/ArtMentor/ArtMentorApp/blob/main/revised_material.md}{link}.
\subsection{HCI Dataset Overview}
\textbf{GPT-4o} \cite{achiam2023gpt} has been selected to assist in art education due to its advanced capabilities in MLLMs. The dataset comprises evaluative feedback from five art teachers on a diverse collection of 20 artworks, covering multiple key dimensions and providing a rich resource for analysis. 
It is worth noting that the teachers and students who contributed to this dataset are from the first and second grades of a primary school, offering a valuable educational perspective from early childhood education.

The dataset includes 20 JSON files documenting the GPT-4o's entity and style recognition results for each artwork. Additionally, there are 360 JSON files that record detailed reviews and suggestions provided by the teachers for each dimension. This extensive dataset, which chronicles 380 sessions, lays a solid foundation for studying the application of MLLMs in art education.
\section{RESULTS ANALYSIS SYSTEM IN \dataset}
In this section, we regard art style as a distinct entity. Consequently, we analyze the results concerning entity classification metrics and art style metrics in Section \ref{ERC}. Next, we examine the score generation capabilities of GPT-4o in Section \ref{SGC}. Finally, we evaluate the review and suggestion capabilities of GPT-4o in Section \ref{RSGC}.
\subsection{Analysis of Entity Recognition Capability}\label{ERC}
To evaluate the entity recognition capabilities of GPT-4o, we \textbf{initially} focus on \textit{art style recognition} as it offers a comprehensive insight into the artwork, forming the basis of our analysis. \textbf{Subsequently}, we examine the \textit{average recognition capabilities} across various entities. \textbf{Finally}, we delve into the finer details within the artwork to assess \textit{nuanced recognition performance}.

\begin{figure}[htbp]
  \centering
   \begin{minipage}[b]{0.49\textwidth}
    \centering
    \includegraphics[height=5cm]{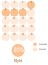}
    \caption{Recognition of art styles by GPT-4o across 20 artworks (Artwork Numbers 1-20).}
    \label{fig:confusion-matrices-part3}
  \end{minipage}
\end{figure}


\begin{figure*}[htbp]
  \centering
  \begin{subfigure}[b]{0.48\textwidth}
    \includegraphics[width=\textwidth]{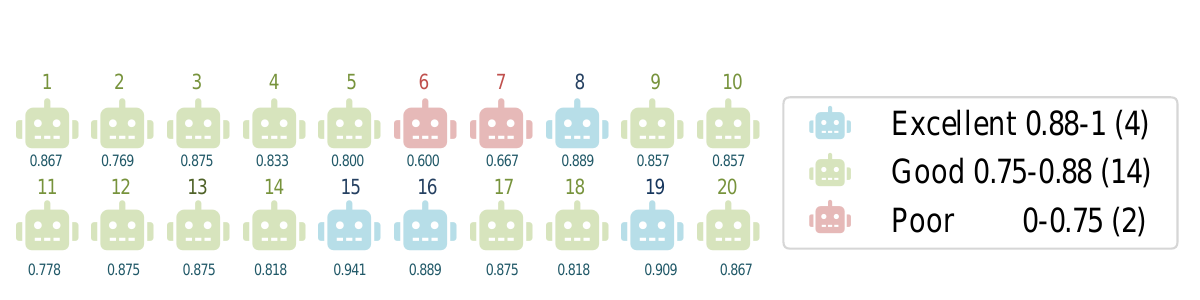}
    \caption{Accuracy.}
    \label{fig:a}
  \end{subfigure}
  \hfill
  \begin{subfigure}[b]{0.50\textwidth}
    \includegraphics[width=\textwidth]{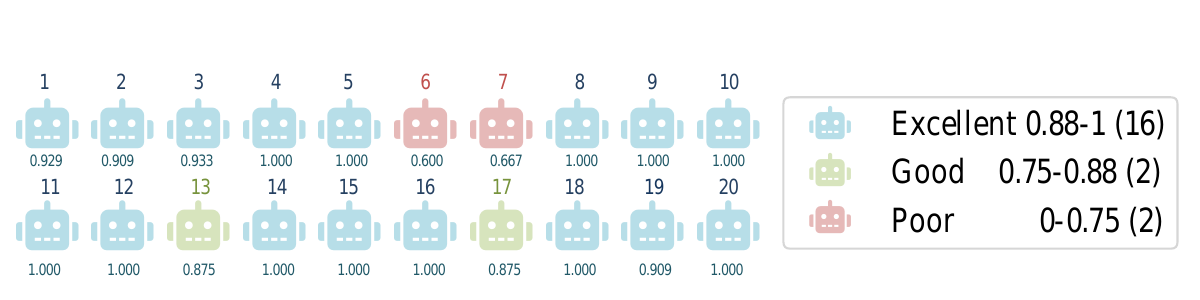}
    \caption{Precision.}
    \label{fig:b}
  \end{subfigure}
  
  \vspace{0.2cm} 

  \begin{subfigure}[b]{0.48\textwidth}
    \includegraphics[width=\textwidth]{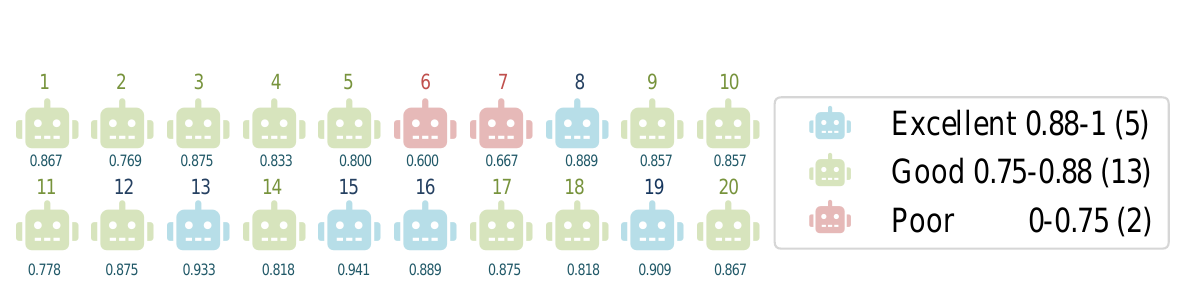}
    \caption{Recall.}
    \label{fig:c}
  \end{subfigure}
  \hfill
  \begin{subfigure}[b]{0.50\textwidth}
    \includegraphics[width=\textwidth]{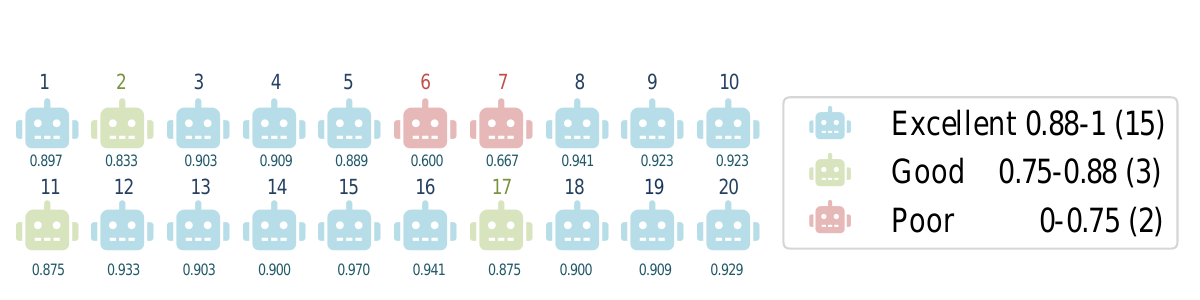}
    \caption{F1-Score.}
    \label{fig:d}
  \end{subfigure}
  
  \caption{Entity classification metrics for GPT-4o across 20 artworks (Artwork Numbers 1-20).}
  \label{fig:confusion-matrices-part1}
\end{figure*}

\subsubsection{How capable is GPT-4o in recognizing art styles?}\label{sec}
As depicted in Figure.~\ref{fig:confusion-matrices-part3}, GPT-4o demonstrates a robust art style recognition capability with an 80\% Art Style Sensitivity (ASS), as detailed in Section \ref{subsec:ass}. Specifically, it accurately identifies narrative illustrations (artworks 1-3) and adheres to the Egyptian frontal law (artworks 8-20).

However, it occasionally misidentifies Chinese ink paintings (artworks 4-7) as watercolors. This error likely arises from a misunderstanding by GPT-4o that Chinese ink paintings can indeed include a variety of colors, not just simply black and white. This misconception underscores the need for model optimization to better understand and differentiate between complex art styles, paving the way for future enhancements.

\begin{figure}[htbp]
  \centering
  \begin{minipage}[b]{0.49\textwidth}
    \centering
    \includegraphics[height=5cm]{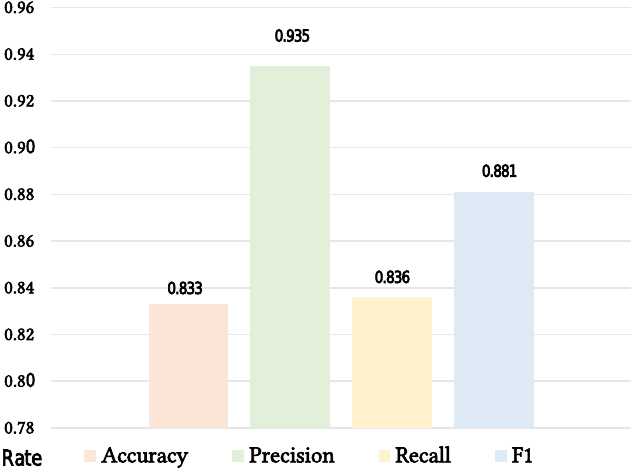}
    \caption{Average entity recognition capability of GPT-4o.}
    \label{fig:confusion-matrices-part2}
  \end{minipage}
\end{figure}

\subsubsection{What is the average entity recognition capability of GPT-4o?} As depicted in Figure.~\ref{fig:confusion-matrices-part2}, precision and F1-score are notably high. The elevated precision may result from a high rate of false negatives (FN), as explained by the precision equation (Eq.~\ref{eq:precision}) and the FN equation (Eq.~\ref{eq:fn}). A detailed examination reveals that $|R_i|$ exceeds $|W_i|$, suggesting art teachers often add more entities than they remove, indicative of a potential bias towards over-recognition. This bias ensures no significant entities are overlooked, aligning with GPT-4o’s strength in minimizing false positives—a key factor in applications that demand high predictive accuracy. Moreover, the F1-score of 0.881 demonstrates a strong, balanced performance in entity recognition between precision and recall.

Interestingly, accuracy (0.833) and recall (0.836) closely align, suggesting minimal false positives (FP), as indicated by the precision formula (Eq.~\ref{eq:precision}) and the false positive formula (Eq.~\ref{eq:fp}). The near equivalence of $|W_i|$ and misrepresentations (MR) supports high precision, showing that art teachers typically replace entities with precision. This behavior underscores the model’s adaptability and conservative approach to entity handling, enhancing its generalization across various art styles.
\subsubsection{How does GPT-4o perform in recognizing artwork details?} To clarify the results further, we illustrate them with a diagram of four waffle charts as shown in Figure. \ref{fig:confusion-matrices-part1}.

The \textbf{first} observation highlights a significant discrepancy between recall (see Figure.~\ref{fig:c}) and accuracy (see Figure.~\ref{fig:a}) in the thirteenth artwork. Unlike other artworks where these metrics are usually similar, this piece features many abstract figures, potentially causing a high number of false positives. The \textbf{second} observation concerns the near absence of false positives. In most artworks, precision is exceptionally high (often 1.000), as shown in Figure.~\ref{fig:b}, suggesting that the model rarely misidentifies non-existent entities and maintains stable performance across various artworks. However, styles like Chinese ink painting may require further optimization to enhance recall. \textbf{Finally}, metrics indicate underperformance in the sixth and seventh artworks, classified under Chinese ink painting. Their abstract nature poses challenges for the model in accurately recognizing and classifying entities.
\subsection{Analysis of Score Generation Capability}\label{SGC}
First, we examine the score volatility of both GPT-4o and art teachers to validate the reliability of score generation. Next, we investigate any significant differences between the initial scores provided by GPT-4o and those assigned by art teachers prior to HCI. Finally, we assess whether the scores from GPT-4o and art teachers converge or diverge after HCI, exploring potential changes towards consensus or further discrepancy. 

\begin{figure*}[htbp]
  \centering
  \begin{subfigure}[b]{0.49\textwidth}
    \includegraphics[width=\textwidth]{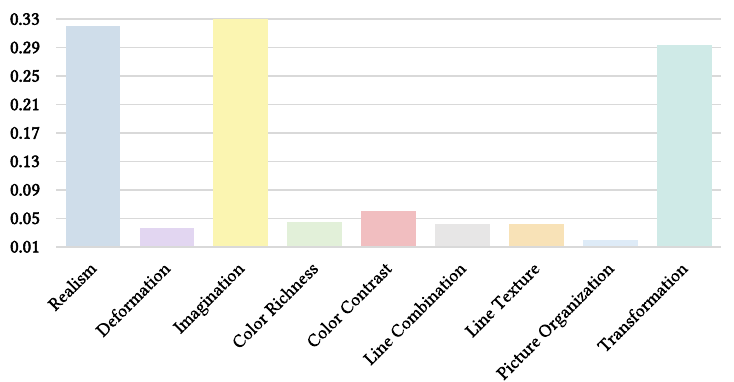}
    \caption{Score Difference (SD).}
    \label{fig:sd}
  \end{subfigure}
  \hfill
  \begin{subfigure}[b]{0.49\textwidth}
    \includegraphics[width=\textwidth]{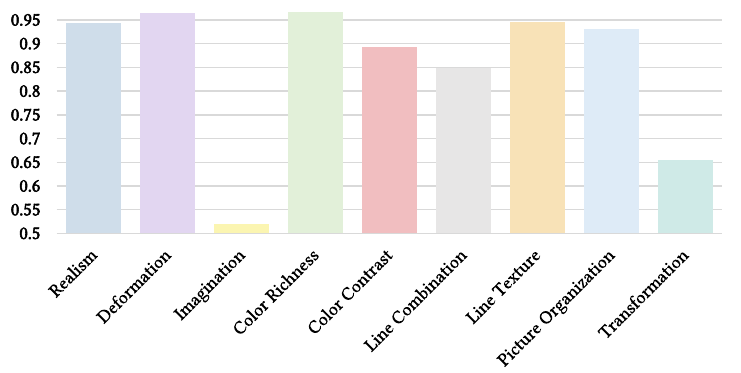}
    \caption{Score Consistency (SC).}
    \label{fig:sc}
  \end{subfigure}
  \caption{Score Acceptance Metrics.}
  \label{fig:sd-sc}
\end{figure*}

\subsubsection{What is the score volatility for GPT-4o and art teachers?} The analysis of Score Volatility (SV) provides significant insights into the stability of scoring by both human scorers and GPT-4o prior to HCI. Art teachers demonstrate consistent scoring patterns, assigning identical scores to the same artwork across similar dimensions throughout the evaluation process. Similarly, GPT-4o, operating at a temperature setting of zero, consistently produces the same outputs for identical artworks under the same conditions. The minimal score volatility observed reinforces the reliability of the dataset, leading us to accept it as robust for further analysis.

\subsubsection{What are the differences between initial scores from GPT-4o and art teachers?}
The Score Difference (SD) metric uncovers scoring discrepancies across various artistic dimensions. As shown in Figure. \ref{fig:sd}, \textbf{\textcolor{art-realism}{Realism}} and \textbf{\textcolor{art-transformation}{Transformation}} display higher SDs of 0.3208 and 0.2941, respectively, highlighting significant divergences between the model's initial assessments and human evaluations.

In contrast, categories such as \textbf{\textcolor{art-deformation}{Deformation}}, \textbf{\textcolor{art-color-richness}{Color Richness}}, \textbf{\textcolor{art-color-contrast}{Color Contrast}}, \textbf{\textcolor{art-line-combination}{Line Combination}}, \textbf{\textcolor{art-line-texture}{Line Texture}}, and \textbf{\textcolor{art-picture-organization}{Picture Organization}} show minimal SDs, indicating a strong concordance between GPT-4o outputs and art teacher adjustments.

\textbf{\textcolor{art-imagination}{Imagination}}, with a moderate SD of 0.5000, signals the potential for further model tuning to achieve closer alignment with expert judgments. Overall, while GPT-4o generally agrees with human scoring, it requires targeted improvements in areas like  \textbf{\textcolor{art-imagination}{Imagination}} to \textit{refine its evaluative accuracy.}
\subsubsection{How do scores from GPT-4o and art teachers change after human-computer interaction?} The Score Consistency (SC) analysis, employing the Spearman correlation coefficient, reveals strong alignment between GPT-4o and art teachers' scores post-HCI, as depicted in Figure. \ref{fig:sc}. SC values for key dimensions are: \textbf{\textcolor{art-realism}{Realism}} (0.9438), \textbf{\textcolor{art-deformation}{Deformation}} (0.9655), \textbf{\textcolor{art-imagination}{Imagination}} (0.5209), and \textbf{\textcolor{art-transformation}{Transformation}} (0.6555), among others. Notably, \textbf{\textcolor{art-realism}{Realism}} saw significant score convergence in later rounds, \textit{suggesting effective model-human integration through iterative feedback.}

However, lower SC values in \textbf{\textcolor{art-imagination}{Imagination}} and \textbf{\textcolor{art-transformation}{Transformation}} highlight areas needing further model calibration to better match human assessments. Overall, the post-HCI data indicates a strong general agreement, with ongoing model refinements crucial for uniform scoring accuracy across all dimensions.
\subsection{Analysis of Review and Suggestion Generation Capability}\label{RSGC}
We evaluate GPT-4o's ability to generate reviews and suggestions using two metrics: the Text Modification Rate (TMR) and Text Similarity (TS). These metrics quantify how well the generated texts meet art teachers' expectations and modifications, providing insights into the effectiveness of the model's text generation. Detailed analyses in the following subsubsections will explore the distinct capabilities of GPT-4o in generating reviews and suggestions.

\subsubsection{How Effective is the Review Generation Capability of GPT-4o?} The Text Modification Rate (TMR) and Text Similarity (TS) metrics provide insights into the alignment of GPT-4o-generated reviews with expert evaluations. 

\begin{figure}[htbp]
  \centering
   \begin{minipage}[b]{0.49\textwidth}
    \centering
    \includegraphics[width=\linewidth]{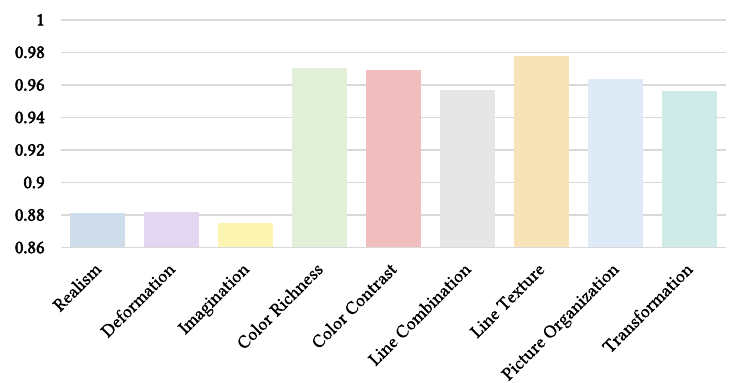}
    \caption{Text Modification Rate (TMR) for R-Agent.}
    \label{fig:tmr-rev}
  \end{minipage}
\end{figure}

As illustrated in Figure. \ref{fig:tmr-rev}, the \textbf{TMR} values, which indicate the extent of textual modifications by art teachers, are shown for various dimensions. These include: \textbf{\textcolor{art-realism}{Realism}} (0.881), \textbf{\textcolor{art-deformation}{Deformation}} (0.882), \textbf{\textcolor{art-imagination}{Imagination}} (0.875), \textbf{\textcolor{art-color-richness}{Color Richness}} (0.971), \textbf{\textcolor{art-color-contrast}{Color Contrast}} (0.969), \textbf{\textcolor{art-line-combination}{Line Combination}} (0.957), \textbf{\textcolor{art-line-texture}{Line Texture}} (0.978), \textbf{\textcolor{art-picture-organization}{Picture Organization}} (0.964), and \textbf{\textcolor{art-transformation}{Transformation}} (0.956). These metrics suggest that the generated reviews were closely aligned with the expectations, requiring only minimal adjustments.

\begin{figure}[htbp]
  \centering
   \begin{minipage}[b]{0.49\textwidth}
    \centering
    \includegraphics[width=\linewidth]{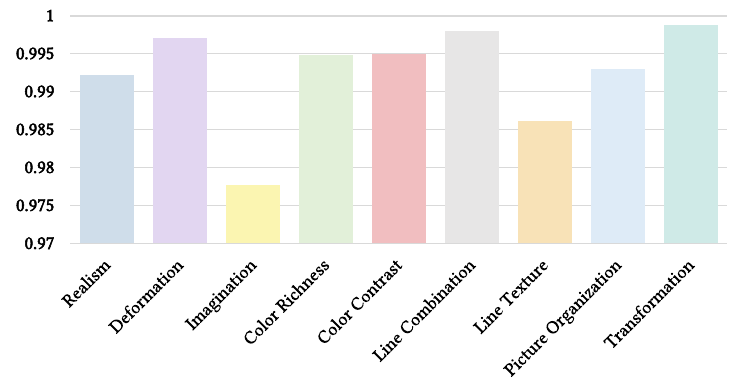}
    \caption{Text Similarity (TS) for R-Agent.}
    \label{fig:ts-rev}
  \end{minipage}
\end{figure}

Additionally, as depicted in Figure. \ref{fig:ts-rev}, the \textbf{TS} metrics highlight the semantic consistency post-modification, with high similarity scores noted across various dimensions: \textbf{\textcolor{art-realism}{Realism}} (0.992), \textbf{\textcolor{art-deformation}{Deformation}} (0.997), \textbf{\textcolor{art-imagination}{Imagination}} (0.978), \textbf{\textcolor{art-color-richness}{Color Richness}} (0.995), \textbf{\textcolor{art-color-contrast}{Color Contrast}} (0.995), \textbf{\textcolor{art-line-combination}{Line Combination}} (0.998), \textbf{\textcolor{art-line-texture}{Line Texture}} (0.986), \textbf{\textcolor{art-picture-organization}{Picture Organization}} (0.993), and \textbf{\textcolor{art-transformation}{Transformation}} (0.999). These scores \textbf{confirm} that \textit{the modifications by art teachers \textbf{maintained} the core content and intent of the original reviews}, verifying GPT-4o’s capability to generate contextually appropriate and stylistically precise content.
\subsubsection{How Effective is the Suggestion Generation Capability of GPT-4o?}
The effectiveness of GPT-4o in generating suggestions is also assessed using the \textbf{TMR} and \textbf{TS} metrics.

\begin{figure}[htbp]
  \centering
   \begin{minipage}[b]{0.49\textwidth}
    \centering
    \includegraphics[width=\linewidth]{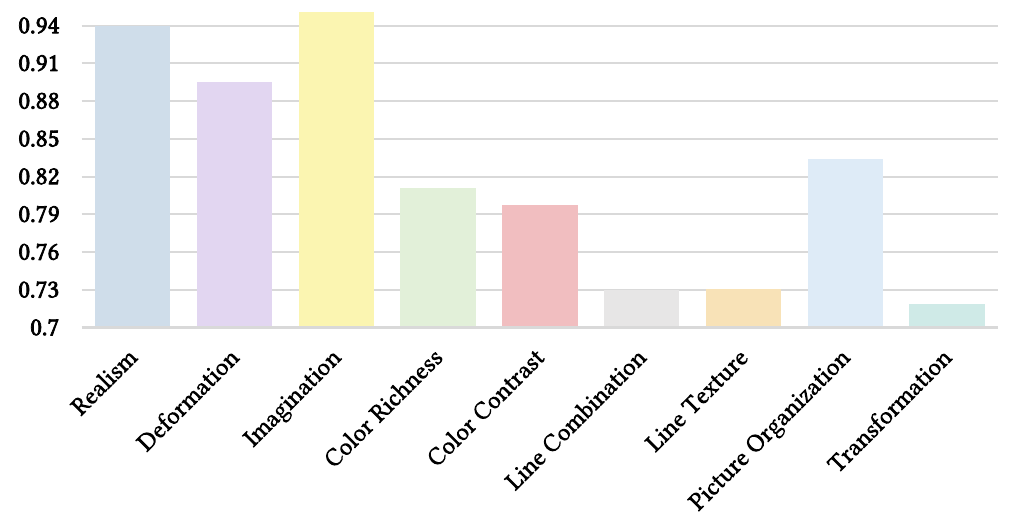}
    \caption{Text Modification Rate (TMR) for S-Agent.}
    \label{fig:tmr-sug}
  \end{minipage}
\end{figure}

As shown in Figure. \ref{fig:tmr-sug}, the \textbf{TMR} values, based on the length of added and deleted content, indicate that GPT-4o performs relatively well in dimensions like \textbf{\textcolor{art-imagination}{Imagination}} (0.968), \textbf{\textcolor{art-realism}{Realism}} (0.940), and \textbf{\textcolor{art-deformation}{Deformation}} (0.895), where fewer modifications were necessary. In contrast, lower TMR values in \textbf{\textcolor{art-transformation}{Transformation}} (0.719), \textbf{\textcolor{art-line-combination}{Line Combination}} (0.730), and \textbf{\textcolor{art-line-texture}{Line Texture}} (0.731) suggest that suggestions in these areas required more extensive revisions by art teachers, both in terms of content addition and deletion.

\begin{figure}[htbp]
  \centering
   \begin{minipage}[b]{0.49\textwidth}
    \centering
    \includegraphics[width=\linewidth]{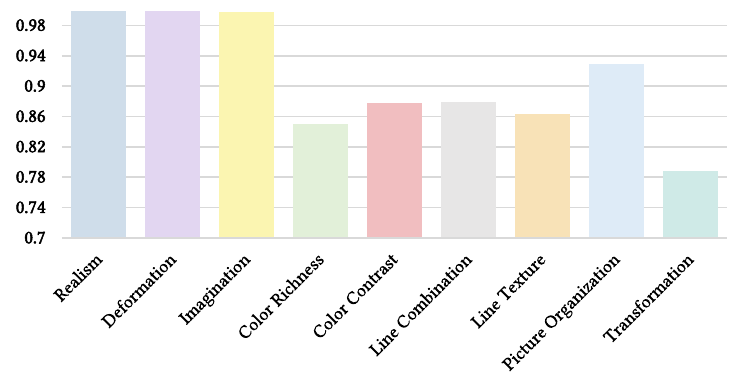}
    \caption{Text Similarity (TS) for S-Agent.}
    \label{fig:ts-sug}
  \end{minipage}
\end{figure}

In Figure. \ref{fig:ts-sug}, the \textbf{TS} values, which focus on word semantics, show high similarity in dimensions such as \textbf{\textcolor{art-realism}{Realism}} (0.999), \textbf{\textcolor{art-deformation}{Deformation}} (0.999), and \textbf{\textcolor{art-imagination}{Imagination}} (0.997), indicating that the core meaning of the suggestions remained largely intact. However, dimensions with lower TMR, such as \textbf{\textcolor{art-transformation}{Transformation}} (0.788), \textbf{\textcolor{art-line-texture}{Line Texture}} (0.863), and \textbf{\textcolor{art-line-combination}{Line Combination}} (0.879), also exhibit lower TS values, suggesting that not only was sentence length altered, but significant semantic changes were made as well. This pattern contrasts with review generation, where modifications affected length without drastically changing semantics, indicating that suggestion generation is less effective than review generation.

Overall, the \textit{dimension requiring the most improvement} is \textbf{\textcolor{art-transformation}{Transformation}}, as it shows substantial modifications in both TMR and TS, reflecting changes in both content length and semantics. Enhancing GPT-4o's suggestion generation in this area would greatly improve its alignment with art teacher expectations.
\section{DISCUSSION}
\label{sec:upgrades}
In this section, we delve into the intriguing question of whether an entity at the intersection of the arts, education, and AI/MLLMs can emerge as an independent assistant within the teacher-student-machine triadic dialogue system for the evaluation of artworks. To address this, we will systematically examine the manifestations of capabilities in multimodal perception, recognition, understanding, and reasoning across the three intelligent agents, thereby shedding light on the potential of such a MLLM to autonomously contribute to artistic assessment.
\subsection{Multimodal Perception and Recognition Capabilities in GPT-4o's Entity Recognition}
In our analysis of entity recognition performance, we observed two key challenges. First, art teachers tended to add more entities than they removed, suggesting that GPT-4o’s initial entity recognition often overlooks relevant details. Second, GPT-4o emphasizes local features, such as facial details or limbs, rather than recognizing complete entities, leading to an over-granulation effect. This reveals a limitation in GPT-4o’s ability to holistically interpret artwork.

However, despite these challenges, GPT-4o demonstrates remarkable multimodal perception and recognition capabilities. The high precision and F1-score across various artworks indicate a strong ability to identify and classify entities accurately. The model’s performance in recognizing art styles, as evidenced by an 80\% Art Style Sensitivity (ASS), further underscores its robustness in understanding and differentiating complex artistic expressions. GPT-4o’s ability to balance between holistic interpretation and detail orientation is noteworthy. While it sometimes misses broader entities, its focus on local features ensures that intricate details are not overlooked. This dual approach is particularly valuable in the context of art analysis, where both the overall style and the finer details contribute to the understanding of the artwork. The model’s adaptability across different art styles, from narrative illustrations to Chinese ink paintings, highlights its generalization capabilities. Although there are areas for improvement, such as better differentiation between Chinese ink paintings and watercolors, the overall performance is commendable. This adaptability is crucial for applications that require consistent performance across a diverse range of artistic styles. The near absence of false positives is a significant strength of GPT-4o. High precision values indicate that the model is conservative in its entity recognition, avoiding the misidentification of non-existent entities. This approach is particularly important in scenarios where accuracy is paramount. While GPT-4o shows excellent performance in both entity and style recognition, there is room for further refinement. Future work could focus on enhancing the model’s ability to recognize broader entities and reducing the over-granulation effect. Additionally, optimizing the model’s understanding of specific art styles, such as Chinese ink painting, could further improve its performance.

In summary, GPT-4o’s entity recognition and style recognition capabilities are excellent. The model’s high precision, strong F1-score, and robust art style sensitivity demonstrate its effectiveness in analyzing artwork. Despite some limitations, the model’s overall performance is impressive, and it holds great potential for future advancements in multimodal perception and recognition.

\subsection{Multimodal Understanding Capability in GPT-4o's Review Generation}
In the \textbf{\textcolor{art-realism}{Realism}} dimension, we observed an encouraging trend toward convergence between GPT-4o and art teachers during the scoring process. Initial discrepancies in assessments gradually diminished as both sides adjusted through repeated interactions, ultimately leading to closer alignment. This suggests potential for improving the model’s adaptability over time.

The convergence observed in the \textbf{\textcolor{art-realism}{Realism}} dimension is a testament to GPT-4o’s ability to learn and adapt through human-computer interaction. This adaptability is crucial for the model’s long-term effectiveness, as it indicates a capacity for continuous improvement in understanding and evaluating artistic nuances. GPT-4o’s performance in generating reviews that align closely with art teachers’ expectations across various dimensions, as evidenced by high TMR and TS metrics, demonstrates its holistic understanding of artwork. The model’s ability to maintain the core content and intent of reviews even after modifications by art teachers highlights its robustness in generating contextually appropriate and stylistically precise content. While the overall performance is strong, a closer look at specific dimensions reveals areas for further refinement. For instance, the \textbf{\textcolor{art-imagination}{Imagination}} dimension shows moderate TMR and TS values, indicating a need for enhanced understanding of abstract and creative aspects of artwork. This suggests that GPT-4o could benefit from additional training data or algorithmic adjustments to better capture the subtleties of imaginative expressions. The iterative feedback loop established between GPT-4o and art teachers plays a pivotal role in refining the model’s review generation capabilities. This loop enables the model to receive targeted feedback, making incremental improvements with each iteration. Continued engagement in this feedback loop is essential for achieving even greater accuracy and alignment with human evaluations. Future work should focus on expanding GPT-4o’s training to include a wider range of artistic styles and dimensions, particularly those that currently show lower convergence rates. Additionally, exploring advanced multimodal understanding techniques, such as integrating visual and textual data more seamlessly, could further enhance the model’s review generation capabilities.

In summary, GPT-4o’s multimodal understanding capability in review generation is robust, with strong performance across various artistic dimensions. The model’s ability to converge with human assessments over time, coupled with its adaptability and holistic review generation, underscores its potential as a valuable tool in art evaluation. Ongoing refinements and iterative feedback will be key to unlocking even greater capabilities in the future.

\subsection{Multimodal Reasoning Capability in GPT-4o's Suggestion Generation}
The analysis of GPT-4o’s suggestion generation capability reveals insights into its multimodal reasoning abilities, particularly in how it integrates visual and textual information to provide constructive feedback.

GPT-4o demonstrates strong performance in certain dimensions, as evidenced by high TMR and TS values in \textbf{\textcolor{art-imagination}{Imagination}}, \textbf{\textcolor{art-realism}{Realism}}, and \textbf{\textcolor{art-deformation}{Deformation}}. These high values indicate that the suggestions generated in these areas were closely aligned with art teachers’ expectations, requiring minimal modifications. This suggests that GPT-4o effectively captures and reasons about the key elements of these artistic dimensions. However, the model faces challenges in more complex dimensions such as \textbf{\textcolor{art-transformation}{Transformation}}, \textbf{\textcolor{art-line-combination}{Line Combination}}, and \textbf{\textcolor{art-line-texture}{Line Texture}}. The lower TMR and TS values in these areas indicate that the suggestions often required significant revisions, both in terms of content length and semantics. This suggests that GPT-4o’s multimodal reasoning capabilities may need further refinement to better understand and articulate the nuances of these dimensions. While GPT-4o shows a strong ability to holistically understand artwork, as seen in its review generation performance, the challenge lies in translating this understanding into detailed, actionable suggestions. The discrepancy between review and suggestion generation effectiveness highlights the need for the model to not only comprehend the artwork but also to provide precise and contextually relevant advice. The iterative feedback loop with art teachers plays a crucial role in enhancing GPT-4o’s suggestion generation capabilities. By receiving detailed feedback on its suggestions, the model can learn to better align its outputs with human expectations. Continued engagement in this feedback loop is essential for refining the model’s multimodal reasoning abilities. To address the challenges identified, future work could explore advanced multimodal reasoning techniques. For instance, integrating more sophisticated visual analysis tools could help the model better understand complex visual elements. Additionally, leveraging techniques such as attention mechanisms and fine-grained feature extraction could enhance the model’s ability to provide detailed and accurate suggestions.

In summary, GPT-4o’s multimodal reasoning capability in suggestion generation shows promise, with strong performance in certain dimensions. However, there is room for improvement, particularly in handling more complex artistic dimensions. Ongoing refinements, iterative feedback, and technological advancements will be key to enhancing the model’s ability to generate contextually appropriate and actionable suggestions, thereby making it an even more valuable tool in the artistic domain.

\section{Conclusion}
In our work, we delve into the fascinating question of whether an entity at the intersection of the arts, education, and AI/MLLMs can emerge as an independent assistant within the teacher-student-machine triadic dialogue system for artwork evaluation. To explore this, we have adopted a Human-Computer Interaction (HCI) space design and analysis approach. We have developed the \dataset~space, which comprises four core components: \textbf{a. Multi-Agent Data Collection System}, \textbf{b. HCI Dataset}, \textbf{c. Data Analysis System}, and \textbf{d. Iterative Upgrades System}. The HCI dataset encompasses 380 sessions across nine dimensions of artwork evaluation, utilizing process-based data to mitigate the inherent manipulation risks associated with outcome-based data. Within the Data Analysis System, we have applied machine learning and natural language processing techniques to imbue the process data with meaning, extracting metrics that objectively reflect MLLM performance while also ensuring interpretability.

Our comprehensive exploration reveals that an entity at the intersection of the arts, education, and AI/MLLMs can indeed emerge as an independent assistant (GPT-4o) within the teacher-student-machine triadic dialogue system for the evaluation of artworks. Despite certain limitations, such as a tendency to overlook broader entities and an over-granulation effect in entity recognition, GPT-4o has demonstrated remarkable capabilities in multimodal perception, recognition, understanding, and reasoning. GPT-4o’s high precision, strong F1-score, and robust art style sensitivity in entity recognition underscore its effectiveness in analyzing artwork. Its ability to adapt and converge with human assessments over time, as evidenced in review generation, highlights its potential for continuous improvement. Furthermore, the model’s performance in generating contextually appropriate and stylistically precise content demonstrates its holistic understanding of artwork. While there are areas for further refinement, particularly in handling more complex artistic dimensions and reducing the over-granulation effect, the overall performance of GPT-4o is impressive. Its adaptability across different art styles and its conservative approach to entity recognition, with a near absence of false positives, are significant strengths. In suggestion generation, GPT-4o shows promise, with strong performance in certain dimensions. However, the challenge lies in translating its holistic understanding into detailed, actionable suggestions.

In summary, GPT-4o holds great potential as a valuable tool in art evaluation within the teacher-student-machine triadic dialogue system. Ongoing refinements, iterative feedback, and technological advancements will be key to unlocking even greater capabilities in the future, thereby further solidifying its role as an independent assistant in the artistic domain.

\bibliographystyle{ACM-Reference-Format}
\bibliography{ref}
\appendix
\section{SYSTEM SETTINGS}\label{ss}
\subsection{Server}
Our system is built using Python and Flask and is deployed locally. The server specifications are as follows:
\begin{itemize}
    \item \textbf{Processor}: Intel Core i7-11800H
    \item \textbf{Memory}: 16GB DDR4 RAM
    \item \textbf{Graphics}: NVIDIA GeForce RTX 3060
    \item \textbf{Storage}: 512GB NVMe SSD
    \item \textbf{Operating System}: Windows 10
\end{itemize}
\subsection{Decoding Parameters}
\label{app:parameters}
We used the following decoding parameters for configuring the inference process of the GPT-4o model:
\begin{itemize}
    \item \textbf{Engine}: GPT-4o
    \item \textbf{Response Length (Word Piece)}: 
    \begin{itemize}
        \item Entity Agent: 100
        \item Review and Suggestion Agents: 500
    \end{itemize}
    \item \textbf{Temperature}: 0 (This parameter controls the randomness of the generated text, with 0 indicating deterministic generation.)
    \item \textbf{Top P}: 1 (This value controls the diversity of candidate words during generation.)
\end{itemize}
\subsection{Prompt Configuration}
When generating outputs for the Entity Agent and Review Agent, we utilized specially designed prompts to guide the model. Below are the details of the prompts for each agent:
\subsubsection{Entity Agent Prompt}
The following prompt was used for the Entity Agent to identify and list objects or features in an image:

\textbf{Prompt:}

\begin{quote}
Identify and list the objects or features present in the image using descriptive labels. Use simple, clear terms like 'Face', 'Black hair', 'Thick lips', 'Big ears', etc. Ensure that each label is descriptive and that labels are separated by the symbol (';'). For example: Face; Black hair; Thick lips; Big ears;. Also, identify the art style of the image with a label starting with '\#\# Style:'.
\end{quote}
\subsubsection{Review Agent Prompts}
The Review Agent generates evaluations based on nine distinct assessment dimensions, with each dimension corresponding to a specific prompt. Below a example of prompts for selected dimensions:

\begin{table}[h]
    \centering
    \caption{Assessment Criteria for Realistic Artwork}
    \noindent
    \begin{tabular}{@{}p{0.1\linewidth}p{0.85\linewidth}@{}}
        \toprule
        \multicolumn{2}{@{}p{\linewidth}@{}}
        {\textbf{Criterion: Realistic. This criterion assesses the accuracy of proportions, textures, lighting, and perspective to create a lifelike depiction.}} \\
        \midrule
        \textbf{5:} & The artwork exhibits exceptional detail and precision in depicting realistic features. Textures and lighting are used masterfully to mimic real-life appearances with accurate proportions and perspective. The representation is strikingly lifelike, demonstrating advanced skills in realism. \\
        \midrule
        \textbf{4:} & The artwork presents a high level of detail and accuracy in the portrayal of subjects. Proportions and textures are very well executed, and the lighting enhances the realism. Although highly realistic, minor discrepancies in perspective or detail might be noticeable. \\
        \midrule
        \textbf{3:} & The artwork represents subjects with a moderate level of realism. Basic proportions are correct, and some textures and lighting effects are used to enhance realism. However, the depiction may lack depth or detail in certain areas. \\
        \midrule
        \textbf{2:} & The artwork attempts realism but struggles with accurate proportions and detailed textures. Lighting and perspective may be inconsistently applied, resulting in a less convincing depiction. \\
        \midrule
        \textbf{1:} & The artwork shows minimal attention to realistic details. Proportions, textures, and lighting are poorly executed, making the depiction far from lifelike. \\
        \bottomrule
    \end{tabular}
\end{table}

\subsubsection{Suggestion Agent Prompt}
The following table outlines the logic used in the Python function that dynamically generates prompts for the Suggestion Agent. This function is designed to provide improvement suggestions for each dimension based on the user-provided data:
\begin{table}[h!]
\centering
\caption{Prompt Generation Logic for Suggestion Agent}
\begin{tabular}{p{2cm} p{6cm}}
\toprule
\textbf{Step} & \textbf{Logic Description} \\ \midrule
Entity Extraction & Extracts entities from the `labels\_data["original"]` list and updates them by removing entities listed in `labels\_data["removed"]`. Newly added entities from `labels\_data["added"]` are inserted into the updated list. The result is concatenated into a string of entity labels separated by semicolons. \\ \midrule
Score Review & Generates a prompt based on the score and review submitted by the user. The current score is extracted from `score\_Review\_data["scores"]["current"]`, and the current review is extracted from `score\_Review\_data["Reviews"]["current"]`. These are combined into a single prompt string. \\ \midrule
Suggestion Inclusion & Incorporates the current suggestion provided by the user into the prompt. The suggestion is extracted from `suggestion\_data["suggestions"]["current"]`. \\ \midrule
Final Prompt Construction & Combines all previous components into the final prompt. The dimension name (`dimension`) is included, and a message is added to inform the Suggestion Agent to consider user feedback and focus on improving the specified dimension. The final prompt instructs the model to output a dictionary format for ease of processing. \\ \bottomrule
\end{tabular}
\end{table}
This function is used to dynamically generate prompts for the Suggestion Agent, providing suggestions for improvements in various dimensions.

For a full list of prompts and detailed information, please refer to the following GitHub repository: \url{https://github.com/ArtMentor/ArtMentorApp/blob/main/ArtMentor_app.py}
\section{INTERVIEW QUESTIONNAIRE}
\subsection*{Greeting}
Dear Participant,

Thank you for taking part in this interview. Your valuable insights will help us further optimize the ArtMentor system, enhancing its role in the field of art education. We are eager to learn about your experiences and suggestions while using the system. Below are some questions we would like to understand from your perspective. We appreciate you taking the time to provide thoughtful responses.

\subsection*{Basic Information}
To better understand your background and how it might influence your interaction with the ArtMentor system, we would appreciate if you could provide the following basic information:
\begin{enumerate}[leftmargin=*]
    \item What is your current role in the field of art education? (e.g., Teacher, Art Mentor, School Administrator, etc.)
    \item How many years of experience do you have in art education?
    \item Have you used other educational technology tools before? If yes, please briefly describe your experience with them.
    \item How often do you integrate technology into your teaching or mentoring activities? (e.g., Daily, Weekly, Occasionally)
    \item What are your main goals when using the ArtMentor system? (e.g., Enhancing student creativity, Providing personalized feedback, etc.)
\end{enumerate}
\subsection*{Main Interview Questions}
\begin{enumerate}[leftmargin=*]
    \item What are your thoughts on the accuracy of the system’s entity recognition in artworks? Do you believe the system met your expectations in this regard?
    \item Regarding the overall usability of the system, how would you rate your experience with the human-computer interaction? If there are any areas for improvement, please provide detailed suggestions.
    \item Do you believe that the multimodal large language model effectively assisted you in scoring, commenting, and generating suggestions for the evaluation of artworks? Please briefly explain your reasoning.
    \item During your use of the system, which functional dimensions  did you find to be particularly well-executed, and which ones do you think require improvement?
    \item Do you consider the suggestions provided by the system to be appropriate for the artistic capabilities of elementary students? Do these suggestions have the potential to inspire creativity and enhance the students' artistic skills?
\end{enumerate}
\subsection*{Closing Remarks}
Once again, thank you for your participation and support! Your feedback will directly contribute to the ongoing improvement of the ArtMentor system, helping us better meet the needs of our users. If you have any additional suggestions or further comments, please do not hesitate to contact us. We look forward to collaborating with you again in the future. Wishing you continued success in your work!
\end{document}